\documentclass[apj,iop]{emulateapj}
 
\usepackage{mathrsfs}
\usepackage{color}
\usepackage{amsbsy}
\usepackage{amsmath}
\usepackage{aas_macros}
\usepackage{url}
\newcommand{\Eq}[1]{Equation (\ref{#1})}
\newcommand{\Sec}[1]{Section~\ref{#1}}
\newcommand{\Fig}[1]{Figure~\ref{#1}}
\newcommand{\Figs}[2]{Figures~\ref{#1} and \ref{#2}}
\newcommand{\Tab}[1]{Table~\ref{#1}}
\newcommand{\Tabs}[2]{Tables~\ref{#1} and \ref{#2}}
\newcommand{\run}[1]{run~\textit{#1}}
\newcommand{\runs}[2]{runs~\textit{#1} and \textit{#2}}
\newcommand{\runss}[3]{runs~\textit{#1}, \textit{#2}, and \textit{#3}}
\newcommand{\runsss}[4]{runs~\textit{#1}, \textit{#2}, \textit{#3}, and
  \textit{#4}} 
\newcommand{\runssss}[5]{runs~\textit{#1}, \textit{#2}, \textit{#3},
  \textit{#4}, and \textit{#5}} 

\newcommand{\vc}[1]{\mbox{\boldmath{$#1$}}}

\newcommand{\dpa}{\partial}
\newcommand{\nab}{\vc{\nabla}}
\newcommand{\St}{\textrm{St}}
\newcommand{\ts}{\tau_{\textrm{f}}}
\newcommand{\Torb}{T_{\textrm{orb}}}
\newcommand{\Msolar}{\mbox{$M_{\normalsize\odot}$}}
\newcommand{\cs}{c_\textrm{s}}

\shorttitle{Planetesimal Formation in Zonal Flows}
\shortauthors{Dittrich, Klahr, \& Johansen}

\begin{document}

\title{GRAVOTURBULENT PLANETESIMAL FORMATION: THE POSITIVE EFFECT OF
  LONG-LIVED ZONAL FLOWS}

\author{K. Dittrich$^1$, H. Klahr$^1$, and A. Johansen$^2$}
\affil{$^1$Max-Planck-Institut f\"ur Astronomie, K\"onigstuhl 17,
  D-69117 Heidelberg, Germany}
\email{dittrich@mpia.de}
\affil{$^2$Lund Observatory, Department of Astronomy and Theoretical Physics,
  Box 43, SE-22100 Lund, Sweden}

\begin{abstract}
  Recent numerical simulations have shown long-lived axisymmetric sub- and
  super-Keplerian flows in protoplanetary disks. These zonal flows are found
  in local as well as global simulations of disks unstable to the
  magnetorotational instability. This paper covers our study of the strength
  and lifetime of zonal flows and the resulting long-lived gas over- and
  underdensities as functions of the azimuthal and radial size of the local
  shearing box. We further investigate dust particle concentrations without
  feedback on the gas and without self-gravity. Strength and lifetime of
  zonal flows increases with the radial extent of the simulation box, but
  decreases with the azimuthal box size. Our simulations support earlier
  results that zonal flows have a natural radial length scale of $5$ to $7$
  gas pressure scale heights. This is the first study that combines
  three-dimensional MHD simulations of zonal flows and dust particles feeling
  the gas pressure. The pressure bumps trap particles with $\St = 1$ very
  efficiently. We show that $\St = 0.1$ particles (of some centimeters in size
  if at $5\,\textrm{AU}$ in an MMSN) reach a hundred-fold higher density than
  initially. This opens the path for particles of $\St = 0.1$ and dust-to-gas
  ratio of $0.01$ or for particles of $\St \geq 0.5$ and dust-to-gas ratio
  $10^{-4}$ to still reach densities that potentially trigger the streaming
  instability and thus gravoturbulent formation of planetesimals.
\end{abstract}

\section{Introduction}\label{intro}

Planets form as a side product in star formation. The general understanding on
how planets in our solar system form was detailed in \citet{S69}. Low-mass
stars form out of molecular clouds which consist of $99\%$ hydrogen and helium
(further referred to as gas), and $1\%$ dust and ices \citep{L03}, i.e.,
everything that has a higher complexity than hydrogen molecules or helium.
Those molecular clouds have cores between less than $0.1$\Msolar and more than
$10$\Msolar \citep{KKM12} that are gravitationally unstable. Most parts of the
mass will collapse into a newborn star. The remaining $\sim 1\%$ of the total
mass will form an accretion disk with pressure supported, sub-Keplerian gas
\citep{W77a,CM81} around the young star. Those disks have lifetimes in the
order of a few million years \citep{HLL01,FAHJO10}. Dust particles grow due to
coagulation \citep{W97}. However, coagulation models show that there are
several barriers to overcome to grow dust large enough to become
gravitationally bound in kilometer-sized planetesimals, such as the bouncing
barrier \citep{ZOGBD10,WBGBDH12,WBOD12}, the fragmentation barrier
\citep[e.g.,][and references therein]{BGBMTW11,BKE12}, and the kilometer-size
barrier \citep{IGM08,CHS08}. Dust growth mechanisms are summarized in
\citet{DBCW07} and the review of \citet{BW08} gives an overview on the
mentioned barriers.

This paper addresses the fragmentation barrier or meter-size barrier. Pebbles
of several decimeters in size will drift very fast inward due to the headwind
from the sub-Keplerian gas \citep{W77a}. Thus, dust has to grow very quickly
from some centimeters to several kilometers in size in order to avoid drifting
into the inner region of the protoplanetary disk.

Turbulence in protoplanetary disks around young stars provides promising
mechanisms for rapid planetesimal formation \citep{JOMKHY07,JKH11}. Shearing
box simulations \citep{BNST95} are a powerful tool for analyzing the
magnetorotational instability \citep[MRI;][]{BH91,BH98} as a source of
turbulence. These simulations consider a local, corotating box, representing
a small part of a Keplerian disk. \citet{JYK09} reported long-lived
axisymmetric sub- and super-Keplerian flows, zonal flows, in shearing box
simulations of turbulence caused by the MRI. These zonal flows have been seen
in several other local \citep{FS09,SG10,SBA12} and global
\citep{LJKP08,DFTKH10,UKFH11,FDKTH11,FDKTH12} simulations using a wide variety
of codes.

Zonal flows are a product of large-scale variations in the magnetic field that
transport momentum differentially, creating regions of slightly faster and
slightly slower rotating gas. Large-scale pressure bumps are excited through
geostrophic balance. This creates long-lived over-densities that potentially
trap dust particles. A more thorough description of zonal flows and their
creation put forward in \citet{JYK09} found zonal flows always populating the
largest radial mode available in the local box approximation. Their largest
box was simulating $10.56$ pressure scale heights ($H$). More recently
\citet{SBA12} found a more complex structure in their largest simulation with
$L_x = 16H$. They further studied the autocorrelation function
\citep{GGSJ09} of the magnetic field and the gas density. Both have a
two-component structure. The first is tilted with respect to the azimuthal
axis and highly localized. The second component is seen at the largest scales
and can be associated with the (predominantly toroidal) background magnetic
field. \citet{SBA12} measure the radial length scale of the zonal flows to
converge at $6H$.

In this paper we consider even larger physical extents for zonal flow
structures. This gives us the opportunity to measure physical properties
such as size and lifetime independent of the simulated domain. Further, we
investigate properties of the zonal flows in radially and azimuthally
stretched boxes. We alter the radial and azimuthal domain up to $\sim\!\!\!20$
gas pressure scale heights.

Additionally, we study the behavior of dust in zonal flows. \citet{W72} was
the first to suggest that axisymmetric pressure bumps can trap
gas. \citet{PBRDUTN12} invoked zonal flows as a possibility to explain the
submillimeter and millimeter-sized particles observed in protoplanetary
disks. They used artificial static density bumps introduced as sinusoidal
density perturbations with different amplitudes (e.g., $A=0.1$ and $A=0.3$)
and different wavelengths ($L = 0.3 \ldots 3 H$). They found that a $30\%$
density perturbation (with $L = 1H$) is necessary to stop the drift of the
dust grains. The present work is the first three-dimensional MHD study that
combines zonal flows and the reaction of dust particles on them.

Our paper is organized as follows. In \Sec{eqs} we discuss the setup of the
simulations in this paper. In \Sec{results1} we study the zonal flow
properties and their dependency on the physical box size. The behavior of dust
particles in zonal flows is described in \Sec{results2}. A discussion and
conclusions follows in \Sec{discussion} and \Sec{summary} provides a summary
and an outlook.

\section{Simulation setup}\label{eqs}

We use the Pencil Code,\footnote{Details on the Pencil Code and download
  information can be found at http://www.nordita.org/software/pencil-code/.} a
sixth-order spatial and third-order temporal finite difference code, for our
simulations. We simulated the standard ideal MHD equations in a local shearing
box with vertical stratification. The simulation boxes are centered at an
arbitrary distance $r$ to the star. The radial direction is denoted by $x$,
the azimuthal direction by $y$, and the vertical direction by $z$. The
Keplerian frequency is $\Omega$. We include dust particle dynamics, without
back-reaction to the gas and without self-gravity.

\subsection{Gas Dynamics}\label{gasdyn}

The gas velocity $\vc{u}$ relative to the Keplerian shear is evolved via the
equation of motion
\begin{eqnarray}  \label{eqnofmotion}
  \frac{\dpa \vc{u}}{\dpa t} &+&  \left( \vc{u} \cdot \nab \right)
  \vc{u} + u_y^{(0)} \frac{\dpa \vc{u}}{\dpa y} = \nonumber \\
  &2& \Omega u_y \hat{\vc{x}} - \frac{1}{2} \Omega u_x \hat{\vc{y}} +
  \Omega^2 z \hat{\vc{z}} \nonumber \\ 
  &+& \frac{1}{\rho} \vc{J} \times \vc{B} - \frac{1}{\rho} \nab P 
  + \vc{f}_\nu \left( \vc{u}, \rho \right) \, .
\end{eqnarray}
On the left-hand side of the equation, the second and third terms are the
advection terms by the perturbed velocity and by shear flow, respectively. The
right-hand side contains the Coriolis force, the vertical component of the
stellar gravity, the Lorentz force, the pressure gradient, and the viscosity
term. Here, $u_y^{(0)} = -(3/2) \Omega x$ is the Keplerian orbital velocity. The
magnetic field $\vc{B}$ as well as the current density $\vc{J}$ are calculated
from the vector potential $\vc{A}$ using $\vc{B} = \nab \times \vc{A}$ and
$\vc{J} = \mu_0^{-1} \nab \times \left( \nab \times \vc{A} \right)$,
respectively. Here, $\mu_0$ is the vacuum permeability. The viscosity term
$\vc{f}_\nu$ is explained in \Sec{visc}.

We evolve the magnetic potential with the uncurled induction equation
\begin{equation}
  \frac{\dpa \vc{A}}{\dpa t} + u_y^{(0)} \frac{\dpa
    \vc{A}}{\dpa y} = \vc{u} \times \vc{B} + \frac{3}{2}
  \Omega A_y \hat{\vc{x}} + \vc{f}_\eta \left( \vc{A} \right)
  \, . \label{inductioneqn} 
\end{equation}
The terms on the right-hand side express the electromotive force, the
stretching (creation of azimuthal magnetic field from radial field) by
Keplerian shear and the resistivity $\vc{f}_\eta$ (see \Sec{resi}).

The gas density is evolved with the continuity equation
\begin{equation}
  \frac{\dpa \rho}{\dpa t} + \left( \vc{u} \cdot \nab \right)
  \rho + u_y^{(0)} \frac{\dpa \rho}{\dpa y} = -\rho \nab \cdot
  \vc{u} + f_D \left( \rho \right) \, , \label{continuityeqn} 
\end{equation}
where the last term on the right-hand side describes mass diffusion (see
\Sec{diff}). We use an isothermal equation of state $P = \cs^2 \rho$, where
the speed of sound is $\cs = H \Omega$; $H$ is the gas pressure scale height.

\subsection{Dissipation} \label{Dissipation}

Maxwell and Reynolds stresses as well as the MRI release kinetic and magnetic
energy at large scales. This energy cascades down to small scales. Since
numerical simulations have a finite resolution, this small-scale energy needs
to be dissipated. We use numerical dissipation in the form of hyper- and shock
viscosity (\Sec{visc}), hyper-resistivity (\Sec{resi}), and hyper- and shock 
diffusion (\Sec{diff}).

\subsubsection{Viscosity} \label{visc}

The viscosity term $\vc{f}_\nu$ in \Eq{eqnofmotion} is expressed by
\begin{eqnarray}\label{eqnviscosity}
  \vc{f}_\nu &=& \nu_3 \left[ \nab^6 \vc{u} + \left( \vc{S}^{(3)} \cdot \nab
      \ln{\rho} \right) \right] \nonumber \\ 
  &+& \nu_{\textrm{sh}} \left[ \nab \nab \cdot \vc{u} + \left( \nab \cdot
      \vc{u} \right) \left( \nab \cdot \ln{\rho} \right) \right]
  \nonumber \\
  &+& \left( \nab \nu_{\textrm{sh}} \right) \nab \cdot \vc{u} \, .
\end{eqnarray}
We restricted our models to hyper-~($\nu_3$) and shock~($\nu_{\textrm{sh}}$)
viscosity. Thus, the regular Navier-Stokes viscosity term is neglected. The
third-order rate-of-strain tensor $\vc{S}^{(3)}$ is defined by
\begin{equation}
  S_{ij}^{(3)} = \frac{ \dpa^5 u_i}{ \dpa x_j^5} \, . \label{eqntensor}
\end{equation}
The high-order Laplacian $\nab^6$ in \Eq{eqnviscosity} is
expanded as $\nab^6 = \dpa^6/\dpa x^6 + \dpa^6/\dpa y^6 + \dpa^6/\dpa
z^6$. Furthermore, the shock viscosity is expressed by
\begin{equation}
  \nu_{\textrm{sh}}=c_{\textrm{sh}} \left< \max{ \left[ - \nab \cdot \vc{u}
      \right]_+} \right> \min{\left( \delta x, \delta y, \delta z \right)}^2
  \, . \label{eqnshockviscosity}
\end{equation}
In the fashion of \citet{vNR50} it is proportional to
positive\footnote{Symbolized by the plus sign in \Eq{eqnshockviscosity}. We
  only apply shock viscosity where the velocity flow is converging.} flow
convergence. We take the maximum over five zones, and smoothed it to the
second order. As suggested by \citet{vNR50}, we set the shock viscosity
coefficient to $c_{sh} = 1.0$ to dissipate energy in shocks at high $z$ above
the mid-plane of the disk.

\subsubsection{Resitivity} \label{resi}

The effects of resistivity are captured by the term
\begin{equation}
  \vc{f}_\eta = \eta_3 \nab^6 \vc{A}
  \, , \label{eqnresistivity}
\end{equation}
where $\eta_3$ is the hyper-resistivity.

\subsubsection{Diffusion} \label{diff}

Mass diffusion is computed with
\begin{equation}
  f_D = D_3 \nab^6 \rho + D_{\textrm{sh}} \nab^2 \rho + \nab D_{\textrm{sh}}
  \cdot \nab \rho \, , \label{eqndiffusion}
\end{equation}
where $D_3$ is the hyper-diffusion parameter and $D_{\textrm{sh}}$ is expanded
as in \Eq{eqnshockviscosity}.

\subsection{Dust Dynamics}\label{dustdyn}

Dust particles are simulated as individual super-particles $i$ with position
$\vc{x}_i$ and velocity $\vc{v}_i$. Each super-particle position is evolved
with 
\begin{equation}
  \frac{d\vc{x}^{(i)}}{dt} = \vc{v}^{(i)} + u_y^{(0)} \hat{\vc{y}}
  \, . \label{eqnofmotionp1} 
\end{equation}
The change of velocity for each particle is evolved through
\begin{eqnarray} \label{eqnofmotionp2}
  \frac{d\vc{v}^{(i)}}{dt} &=& 2 \Omega v_y^{(i)} \hat{\vc{x}} -
  \frac{1}{2} \Omega v_x^{(i)} \hat{\vc{y}} - \Omega^2 z \hat{\vc{z}}
  \nonumber \\ 
  &-& \frac{1}{\ts} \left[ \vc{v}^{(i)} - \vc{u} (\vc{x}^{(i)})
    \right] \, ,
\end{eqnarray}
where the first and second terms are due to the Coriolis force. The third term
corresponds to the vertical gravity of the star. Particles only feel the gas
drag (the last term in \Eq{eqnofmotionp2}) of nearby cells, but are
not subjected to pressure or Lorentz forces. $\ts$ denotes the friction
time, a measure for the size of the particles.

\subsection{Boundary Conditions}\label{boundaryconditions}

For our simulations, we use shearing box boundary conditions in radial
(shear-periodic) and azimuthal (periodic) directions. In the vertical
direction we also use periodic boundary conditions. Although periodic boundary
conditions in vertical direction are not physical, these boundary conditions
conserve the average flux of the magnetic field. Simulations with outflow
boundaries (not included in this paper) showed no considerable mass flux
across the vertical boundary and did not change the average properties of the
zonal flow.

\subsection{Dimensions}\label{units}

We use the dimensionless unit system $\cs = \Omega = \mu_0 = \rho_0 =
1$. Velocity is measured in units of the local sound speed $\cs$. Gas
velocities are always denoted by $\vc{u}$ whereas particle velocities are
always denoted by $\vc{v}$. All velocities are differences to the Keplerian
orbital velocity $\vc{v}_K = (0, u_y^{(0)}, 0)$, where $u_y^{(0)} =
-(3/2) \Omega x$. Time is measured in units of the local orbital time
$\Torb = 2 \pi \Omega^{-1}$. Length measures are in units of the
pressure scale height $H = \cs \Omega^{-1}$. Density is stated in units of the
initial mid-plane gas density $\rho_0$. Magnetic field strength is measured in
units of $\cs (\mu_0 \rho_0)^{-1}$. Energy and stress are in units of the mean
thermal pressure in the box $\left< P \right> = \cs^2 \left< \rho \right>$.

Since our simulations are dimensionless, they can be placed at any distance
$r$ to the star. Only by defining a global pressure gradient $\dpa P_{\textrm
  global} / \dpa r$, which balances the Coriolis force in
\begin{equation}
    \frac{1}{\rho} \frac{\dpa P_{\textrm{global}}}{\dpa r} = 2 \Omega \Delta v
    \, , \label{globaldpdr}
\end{equation}
we restrict our simulations to a specific distance to the star where the
chosen pressure gradient applies. The parameter $\Delta v = u_y^{(0)} -
u_y$ is the difference to the azimuthal Keplerian velocity. We fix $\Delta v =
0.05 \cs$ (see also \Sec{initialcond}). Numerically, the global pressure
gradient acts as an external force on gas and dust.

\subsection{Initial Conditions}\label{initialcond}

The gas density is set to an isothermal hydrostatic equilibrium
$\rho(z)=\rho_0 \exp{(-z^2/2H^2)}$. We start with random noise fluctuations in
the gas velocity with $\delta \vc{u} = 10^{-3} \cs$. The azimuthal component
of the magnetic vector potential is initialized with $A_y = A_0 \cos{(k_x x)}
\cos{(k_y y)} \cos{(k_z z)}$ where throughout $k_x = k_y = k_z = 4.76 H^{-1}$
and $A_0 = 0.04 \cs(\mu_0 \rho_0)^{-1}$.

Particles are released after the gas turbulence is saturated. We measured this
to be after $20 \Torb$ for the largest runs. For convenience, we used the
same saturation time for all our simulations. Particles have a Stokes number
of $\St = \ts \Omega = 1$, unless otherwise stated. The initial particle
distribution is Gaussian in $z$ and uniform in $x$ and $y$. The particle
velocity is initialized with the stationary solution \citep{NSH86} for the
radial and azimuthal velocity
\begin{eqnarray} \label{partvelinit}
  \frac{v_x}{\cs} &=& - \frac{2 \Delta v}{\ts \Omega + (\ts \Omega)^{-1}}
  \nonumber \\
  \frac{v_y}{\cs} &=& - \frac{\Delta v}{1 + (\ts \Omega)^2} \, .
\end{eqnarray}
We get $\Delta v$ from the solution of \Eq{globaldpdr}
\begin{equation}
  \frac{\Delta v}{\cs} = - \frac{1}{2} \left( \frac{H}{r} \right)^2 \frac{\dpa
    \ln{P}}{\dpa \ln{r}} \, . \label{deltavsolved}
\end{equation}
We initialized $\Delta v = 0.05 \cs$ for our simulations.

\subsection{Simulation Parameters}\label{simulationparameters}

\begin{table*}[t!]
  \centering
    \caption{Run Parameters} \label{runparameters}
    \begin{tabular}{clcclrccl}
      \hline \hline
      Simulation Set & Run & $L_x \times L_y \times L_z$ & $N_x \times N_y
      \times N_z$ & $\nu_3=\eta_3=D_3$ & $n_\textrm{particles}$ & $\St$ &
      Shear & $\Delta t$ \\
      \multicolumn{1}{c}{(1)} & \multicolumn{1}{c}{(2)} &
      \multicolumn{1}{c}{(3)} & \multicolumn{1}{c}{(4)} &
      \multicolumn{1}{c}{(5)} & \multicolumn{1}{c}{(6)} &
      \multicolumn{1}{c}{(7)} & \multicolumn{1}{c}{(8)} & 
      \multicolumn{1}{c}{(9)} \\
      \hline
      \textit{A}     & S        & $ 1.32 \times  1.32 \times 2.64$ & 
      $ 36 \times  36 \times 72$ & $4.0 \times 10^{-10}$ &
         $62$,$500$ & $1.0$ & FDA & $121$ \\ 
      \textit{A,B,C} & M        & $ 2.64 \times  2.64 \times 2.64$ & 
      $ 72 \times  72 \times 72$ & $4.0 \times 10^{-10}$ & 
        $250$,$000$ & $1.0$ & FDA & $121$ \\
      \textit{A,E}   & L        & $ 5.28 \times  5.28 \times 2.64$ &
      $144 \times 144 \times 72$ & $4.0 \times 10^{-10}$ & 
      $1$,$000$,$000$ & $1.0$ & FDA & $121$ \\ 
      \textit{A,E}   & XL       & $10.56 \times 10.56 \times 2.64$ &
      $288 \times 288 \times 72$ & $4.0 \times 10^{-10}$ &
      $4$,$000$,$000$ & $1.0$ & FDA & $121$ \\
      \textit{A}     & XXL      & $21.12 \times 21.12 \times 2.64$ &
      $576 \times 576 \times 72$ & $4.0 \times 10^{-10}$ &
      $4$,$000$,$000$ & $1.0$ & FDA & $121$ \vspace{0.1cm} \\
      
      \textit{B}     & x-S      & $ 1.32 \times  2.64 \times 2.64$ &
      $ 36 \times  72 \times 72$ & $4.0 \times 10^{-10}$ &
        $125$,$000$ & $1.0$ & FDA & $121$ \\
      \textit{B}     & x-L      & $ 5.28 \times  2.64 \times 2.64$ &
      $144 \times  72 \times 72$ & $4.0 \times 10^{-10}$ &
        $500$,$000$ & $1.0$ & FDA & $121$ \\
      \textit{B}     & x-XL     & $10.56 \times  2.64 \times 2.64$ &
      $288 \times  72 \times 72$ & $4.0 \times 10^{-10}$ & 
      $1$,$000$,$000$ & $1.0$ & FDA & $121$ \vspace{0.1cm} \\

      \textit{C}     & y-S      & $ 2.64 \times  1.32 \times 2.64$ &
      $ 72 \times  36 \times 72$ & $4.0 \times 10^{-10}$ &
        $125$,$000$ & $1.0$ & FDA & $121$ \\
      \textit{C}     & y-L      & $ 2.64 \times  5.28 \times 2.64$ &
      $ 72 \times 144 \times 72$ & $4.0 \times 10^{-10}$ & 
        $500$,$000$ & $1.0$ & FDA & $121$ \\
      \textit{C}     & y-XL     & $ 2.64 \times 10.56 \times 2.64$ & 
      $ 72 \times 288 \times 72$ & $4.0 \times 10^{-10}$ & 
      $1$,$000$,$000$ & $1.0$ & FDA & $121$ \vspace{0.1cm} \\

      \textit{D}     & LspecMR  & $ 5.28 \times  5.28 \times 2.64$ & 
      $144 \times 144 \times 72$ & $4.0 \times 10^{-10}$ &
      $1$,$200$,$000$ & $0.01 \ldots 100$ & FDA & $121$ \\ 
      \textit{D}     & LspecHR  & $ 5.28 \times  5.28 \times 2.64$ & 
      $256 \times 256 \times 128$ & $2.0 \times 10^{-11}$ & 
      $120$,$000$,$000$ & $0.01 \ldots 100$ & FDA & $121$ \\ 
      \textit{D}     & LspecMRs & $ 5.28 \times  5.28 \times 2.64$ & 
      $144 \times 144 \times 72$ & $4.0 \times 10^{-10}$ &
      $14$,$000$,$000$ & $0.01 \ldots 1.0$ & FDA & $121$ \\ 
      \textit{D}     & MspecMRb & $ 2.64 \times  2.64 \times 2.64$ & 
      $ 72 \times  72 \times 72$ & $4.0 \times 10^{-10}$ &
      $6$,$000$,$000$ & $1.0 \ldots 100$ & FDA & $223$ \vspace{0.1cm} \\

      \textit{E}     & L\_SAFI  & $ 5.28 \times  5.28 \times 2.64$ &
      $144 \times 144 \times 72$ & $4.0 \times 10^{-10}$ & 
      $1$,$000$,$000$ & $1.0$ & SAFI & $121$ \\ 
      \textit{E}     & XL\_SAFI & $10.56 \times 10.56 \times 2.64$ &
      $288 \times 288 \times 72$ & $4.0 \times 10^{-10}$ &
      $4$,$000$,$000$ & $1.0$ & SAFI & $121$ \\
      \hline
    \end{tabular}
  {\textbf Notes.} Column 1: simulation set. Column 2: name of run. Column
  3: box size in units of pressure scale heights. Column 4: grid
  resolution. Column 5: dissipation coefficients. Column 6: number of
  particles in simulation. Column 7: Stokes number $\St = \ts \Omega$.
  Column 8: shear advection scheme. Column 9: total run time in orbits
  $\Torb$.
\end{table*}

\begin{figure}[htb]
  \centering
    \includegraphics[width=\linewidth]{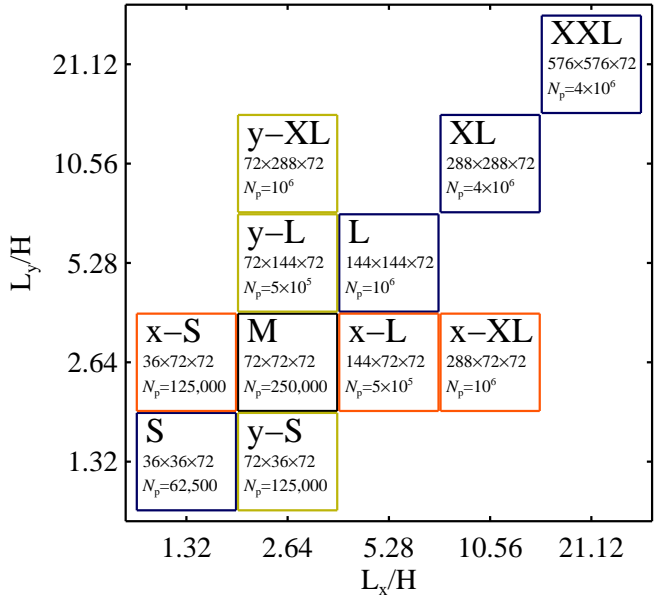}
  \caption{Parameter space of radial and azimuthal box sizes that was
    simulated for this paper. Every simulation has a vertical extent of
    $2.64H$. The first line in each box states the name of the run, the second
    line the number of grid cells used, and the third line gives the number of
    simulated super-particles. More details on all simulations are found in
    \Tab{runparameters}.}
  \label{parspc}
\end{figure}

The parameter space covered by our simulations is summarized in
\Fig{parspc}. The vertical extent is always set to
$L_z=2.64H$.\footnote{$L=1.32$ has been chosen as the basic box size,
  because $L_x=1.32$ approximately marks the transition from subsonic
  to supersonic Keplerian shear flow \citep{JYK09}.} One simulation
set~(\textit{A}) covers the boxes with a squared base, i.e., radial and
azimuthal extent are kept the same:
$L_x~=~L_y~=~\left\{1.32,~2.64,~5.28,~10.56,~21.12\right\}H$. These
are marked with blue boxes in \Fig{parspc} and are called
\runssss{S}{M}{L}{XL}{XXL}. The deviation to the global density profile in the
largest box can be quite severe at the inner and outer boundary of the largest
simulation. Thus, the results from \run{XXL} have to be treated with
caution. Another set of simulations~(\textit{B}) varies the radial size of the
box, $L_x~=~\left\{1.32,~2.64,~5.28,~10.56\right\}H$, with constant box size
in azimuthal direction, $L_y=2.64H$. This set is marked red in \Fig{parspc}
and includes \runsss{x-S}{M}{x-L}{x-XL}. The third set of
simulations~(\textit{C}) varies the azimuthal extent,
$L_y~=~\left\{1.32,~2.64,~5.28,~10.56\right\}H$, while the radial 
extent is kept constant, $L_x=2.64H$. This set includes
\runsss{y-S}{M}{y-L}{y-XL} (marked yellow in \Fig{parspc}). All simulations
are stratified and have dust particles with different couplings to the
gas. The simulations displayed in \Fig{parspc} have particles with a Stokes
number of $\St = 1$.

Details on run parameters of those and six more simulations are found in
\Tab{runparameters}. The first set of simulations (\textit{A},\textit{B}, and
\textit{C}) in \Tab{runparameters} are the simulations with medium resolution,
i.e., $36$ grid cells\footnote{We chose $36$ grid cells instead of the usual
  $32$ grid cells. That choice was done due to the architecture (12 CPUs per
  node) of the used cluster, THEO in the MPG computing center in Garching.}
per $1.32$ pressure scale heights. Simulation set \textit{D} was carried out
to investigate the behavior of different particle sizes in the presence of
zonal flows. Run \textit{LspecMR} is very much like \run{L}, but with $12$
different particle Stokes numbers. The \run{LspecHR} has a resolution of $64$
grid cells per $1.32H$. Runs \textit{LspecMR} and \textit{LspecHR} have $12$
different particle species, with Stokes numbers of $\St = 0.01 \ldots
100.0$. The \runs{LspecMRs}{MspecMRb} have particles with Stokes numbers of
$\St = 0.01 \ldots 1.0$ and $\St = 1.0 \ldots 100.0$, respectively. These two
simulations were carried out to study particle behavior with more particles
per grid cell\footnote{$\sim\!9$ and $\sim\!16$ particles per grid cell for
  \runs{LspecMRs}{MspecMRb} compared to $\sim\!0.8$ particles per grid cell
  for \run{LspecMR}.} at medium resolution. The corresponding sizes for
different protoplanetary disk models are found in \Sec{dustdisc}.

Simulation set \textit{E} is a comparison of \runs{L}{XL} to the same runs
(\textit{L\_SAFI} and \textit{XL\_SAFI}) with the Shear Advection by Fourier 
Interpolation (SAFI) scheme. Here, all variables $q(x, y, z)$ are transformed
into Fourier space in the $y$-direction to get $\hat{q}(x, k_y, z)$. Then each
Fourier mode is multiplied by $\exp{[\textrm{i} k_y u_y^{(0)}(x) \delta t]}$
to shift by $u_y^{(0)}(x) \delta t$ in real space and is inverse Fourier
transformed to real space. This method reduces the advection error to the
standard Finite Difference Advection (FDA) scheme in the Pencil Code
\citep[more details on FDA and SAFI are found in][]{JYK09}.

\begin{table*}[htb]
  \centering
    \caption{Turbulence Properties} \label{turbresults}
    \begin{tabular}{lccccccccc}
      \hline \hline
      Run &
      $\langle \tfrac{1}{2} u_x^2 \rangle$ &
      $\langle \tfrac{1}{2} u_y^2 \rangle$ &
      $\langle \tfrac{1}{2} u_z^2 \rangle$ &
      $\langle \tfrac{1}{2} B_x^2 \rangle$ &
      $\langle \tfrac{1}{2} B_y^2 \rangle$ &
      $\langle \tfrac{1}{2} B_z^2 \rangle$ &
      $\langle \rho u_xu_y \rangle$ &
      $\langle -B_xB_y \rangle$ &
      $\alpha$ \\
      \multicolumn{1}{c}{(1)} & \multicolumn{1}{c}{(2)} &
      \multicolumn{1}{c}{(3)} & \multicolumn{1}{c}{(4)} &
      \multicolumn{1}{c}{(5)} & \multicolumn{1}{c}{(6)} &
      \multicolumn{1}{c}{(7)} & \multicolumn{1}{c}{(8)} &
      \multicolumn{1}{c}{(9)} & \multicolumn{1}{c}{(10)}\\
      \hline
      S &
      $2.1 \times 10^{-3}$ & $3.3 \times 10^{-3}$ &
      $1.5 \times 10^{-3}$ & 
      $8.8 \times 10^{-4}$ & $6.4 \times 10^{-3}$ &
      $3.5 \times 10^{-4}$ &
      $7.7 \times 10^{-4}$ & $3.4 \times 10^{-3}$ &
      $2.8 \times 10^{-3}$ \\
      M &
      $3.9 \times 10^{-3}$ & $5.2 \times 10^{-3}$ &
      $2.2 \times 10^{-3}$ & 
      $1.9 \times 10^{-3}$ & $1.2 \times 10^{-2}$ &
      $7.6 \times 10^{-4}$ &
      $1.6 \times 10^{-3}$ & $6.6 \times 10^{-3}$ &
      $5.5 \times 10^{-3}$ \\
      L &
      $5.0 \times 10^{-3}$ & $5.6 \times 10^{-3}$ &
      $2.3 \times 10^{-3}$ & 
      $2.0 \times 10^{-3}$ & $1.3 \times 10^{-2}$ &
      $8.0 \times 10^{-4}$ &
      $1.9 \times 10^{-3}$ & $6.9 \times 10^{-3}$ &
      $5.9 \times 10^{-3}$ \\
      XL &
      $5.2 \times 10^{-3}$ & $5.0 \times 10^{-3}$ &
      $2.1 \times 10^{-3}$ & 
      $1.7 \times 10^{-3}$ & $1.1 \times 10^{-2}$ &
      $6.7 \times 10^{-4}$ &
      $1.8 \times 10^{-3}$ & $6.1 \times 10^{-3}$ &
      $5.2 \times 10^{-3}$ \\
      XXL &
      $5.1 \times 10^{-3}$ & $4.6 \times 10^{-3}$ &
      $2.0 \times 10^{-3}$ & 
      $1.6 \times 10^{-3}$ & $1.0 \times 10^{-2}$ &
      $6.2 \times 10^{-4}$ &
      $1.7 \times 10^{-3}$ & $5.7 \times 10^{-3}$ &
      $4.9 \times 10^{-3}$ \vspace{0.1cm} \\

      x-S &
      $4.0 \times 10^{-3}$ & $5.2 \times 10^{-3}$ &
      $2.4 \times 10^{-3}$ & 
      $2.0 \times 10^{-3}$ & $1.3 \times 10^{-2}$ &
      $8.4 \times 10^{-4}$ &
      $1.7 \times 10^{-3}$ & $7.0 \times 10^{-3}$ &
      $5.8 \times 10^{-3}$ \\
      x-L &
      $3.8 \times 10^{-3}$ & $5.2 \times 10^{-3}$ &
      $2.1 \times 10^{-3}$ & 
      $1.7 \times 10^{-3}$ & $1.1 \times 10^{-2}$ &
      $6.9 \times 10^{-4}$ &
      $1.5 \times 10^{-3}$ & $6.1 \times 10^{-3}$ &
      $5.1 \times 10^{-3}$ \\
      x-XL &
      $3.8 \times 10^{-3}$ & $5.0 \times 10^{-3}$ &
      $2.2 \times 10^{-3}$ & 
      $1.8 \times 10^{-3}$ & $1.2 \times 10^{-2}$ &
      $7.1 \times 10^{-4}$ &
      $1.5 \times 10^{-3}$ & $6.2 \times 10^{-3}$ &
      $5.2 \times 10^{-3}$ \vspace{0.1cm} \\

      y-S &
      $2.3 \times 10^{-3}$ & $3.7 \times 10^{-3}$ &
      $1.6 \times 10^{-3}$ & 
      $1.0 \times 10^{-3}$ & $7.1 \times 10^{-3}$ &
      $4.1 \times 10^{-4}$ &
      $8.6 \times 10^{-4}$ & $3.8 \times 10^{-3}$ &
      $3.1 \times 10^{-3}$ \\
      y-L &
      $5.2 \times 10^{-3}$ & $5.7 \times 10^{-3}$ &
      $2.5 \times 10^{-3}$ & 
      $2.2 \times 10^{-3}$ & $1.4 \times 10^{-2}$ &
      $8.8 \times 10^{-4}$ &
      $2.0 \times 10^{-3}$ & $7.5 \times 10^{-3}$ &
      $6.3 \times 10^{-3}$ \\
      y-XL &
      $5.4 \times 10^{-3}$ & $5.0 \times 10^{-3}$ &
      $2.2 \times 10^{-3}$ & 
      $1.8 \times 10^{-3}$ & $1.2 \times 10^{-2}$ &
      $7.2 \times 10^{-4}$ &
      $1.9 \times 10^{-3}$ & $6.4 \times 10^{-3}$ &
      $5.6 \times 10^{-3}$ \vspace{0.1cm} \\

      LspecMR &
      $4.8 \times 10^{-3}$ & $5.3 \times 10^{-3}$ &
      $2.2 \times 10^{-3}$ & 
      $1.9 \times 10^{-3}$ & $1.2 \times 10^{-2}$ &
      $7.5 \times 10^{-4}$ &
      $1.8 \times 10^{-3}$ & $6.6 \times 10^{-3}$ &
      $5.6 \times 10^{-3}$ \\
      LspecHR &
      $3.0 \times 10^{-3}$ & $4.2 \times 10^{-3}$ &
      $1.4 \times 10^{-3}$ & 
      $1.3 \times 10^{-3}$ & $7.5 \times 10^{-3}$ &
      $5.4 \times 10^{-4}$ &
      $1.0 \times 10^{-3}$ & $4.2 \times 10^{-3}$ &
      $3.5 \times 10^{-3}$ \\
      LspecMRs &
      $5.3 \times 10^{-3}$ & $6.2 \times 10^{-3}$ &
      $2.5 \times 10^{-3}$ & 
      $2.2 \times 10^{-3}$ & $1.4 \times 10^{-2}$ &
      $9.2 \times 10^{-4}$ &
      $2.1 \times 10^{-3}$ & $7.7 \times 10^{-3}$ &
      $6.5 \times 10^{-3}$ \\
      MspecMRb &
      $4.3 \times 10^{-3}$ & $5.7 \times 10^{-3}$ &
      $2.5 \times 10^{-3}$ & 
      $2.1 \times 10^{-3}$ & $1.4 \times 10^{-2}$ &
      $8.7 \times 10^{-4}$ &
      $1.8 \times 10^{-3}$ & $7.2 \times 10^{-3}$ &
      $6.0 \times 10^{-3}$ \vspace{0.1cm} \\

      L\_SAFI &
      $5.1 \times 10^{-3}$ & $5.8 \times 10^{-3}$ &
      $2.4 \times 10^{-3}$ & 
      $2.1 \times 10^{-3}$ & $1.3 \times 10^{-2}$ &
      $8.4 \times 10^{-4}$ &
      $2.0 \times 10^{-3}$ & $7.2 \times 10^{-3}$ &
      $6.1 \times 10^{-3}$ \\
      XL\_SAFI &
      $5.4 \times 10^{-3}$ & $5.3 \times 10^{-3}$ &
      $2.2 \times 10^{-3}$ & 
      $1.8 \times 10^{-3}$ & $1.2 \times 10^{-2}$ &
      $7.3 \times 10^{-4}$ &
      $1.9 \times 10^{-3}$ & $6.5 \times 10^{-3}$ &
      $5.6 \times 10^{-3}$ \\
      \hline
    \end{tabular}
  {\textbf Notes.} Column 1: name of run. Columns 2-4: kinetic energy. Columns
  5-7: magnetic energy. Column 8: Reynolds stress. Column 9: Maxwell
  stress. Column 10: $\alpha$-value, following \Eq{eqnalpha}. Stresses and
  energies have been normalized to the mean thermal pressure in the box,
  $\langle P \rangle = \cs^2 \langle \rho \rangle$. 
\end{table*}

Every simulation is run for $121$ local orbits $\Torb = 2 \pi \Omega^{-1}$,
except for \textit{LspecMRb} which runs for $223 \Torb$ in order to follow the
evolution of the slowly settling large particles. After $20\Torb$, when the
initial conditions are sufficiently forgotten and the turbulence saturated,
the particles are started.

\section{Zonal Flow Properties}\label{results1} 

Turbulence properties are summarized in \Tab{turbresults}. The kinetic and
magnetic energy as well as the Reynolds and Maxwell stress almost doubles
when increasing the box size from $(1.32H)^2 \times 2.64H$ (\run{S}) to
$(2.64H)^3$ (\run{M}). Further increasing the box size does not change the
resulting energies and stresses by much. The radially short box of
\run{x-S} with $1.32H \times (2.64H)^2$ has similar results on these values.
However, the azimuthally short box of \run{y-S} has turbulent energies and
stresses comparable to \run{S}. These measurements show that the turbulence
parameters are saturated for boxes with an azimuthal extent of at least
$2.64H$. This confirms the results from \citet{FS09} who found that the
turbulence properties do not change when the box size is increased radially,
if the azimuthal dimension is large enough. The $\alpha$-value \citep{SS73} in
Column 10 in \Tab{turbresults} is calculated via 
\begin{equation}
  \alpha = \frac{2}{3} \frac{\left( \left< \rho u_x u_y \right> - \left< B_x
        B_y \right> \right)}{\left< P \right>} \, , \label{eqnalpha}
\end{equation}
where $\left< P \right> = \cs^2 \left< \rho \right>$. The factor of $2/3$
originates from the shear parameter $q = -d \ln{\Omega}/d\ln{R}$. We use
$q = 3/2$, appropriate for a Keplerian disk. For further details see
\citet[page 748]{BNST95}. The Maxwell stress is around three times higher than
the Reynolds stress and thus dominates the $\alpha$-value.

In order to verify that our numerical resolution is sufficient, we examined
the quality factor as described in \citet{SBA12}:
\begin{equation} \label{eqnqfactor}
  Q_j = \frac{\overline{2 \pi |v_\textrm{a,j}|}}{\Omega \Delta x_j} \, ,
\end{equation}
where the Alfv\'{e}n speed is defined as $|v_\textrm{a,j}|^2 = \langle B_j
\rangle^2 / \langle \rho \rangle$. The notation $\langle x \rangle$ denotes
volume averaging, $\overline{x}$ shows a time average. \citet{SRSB12} show
that $Q_z \gtrsim 10-15$ for poorly resolved azimuthal quality factors ($Q_y
\sim 10$) is required to resolve the MRI. Larger values of the azimuthal
quality factor ($Q_y \gtrsim 25$) allow for lower vertical quality
factors. The azimuthal component of the magnetic field is very well resolved
($Q_y \gtrsim 25$) for all simulations, but \runs{S}{y-S}. The vertical
component has values between $6$ and $8$. We thus conclude that all
simulations, but \runs{S}{y-S} have sufficient resolution for the MRI.

\begin{figure*}[htb]
  \centering{\includegraphics[width=0.65\linewidth]{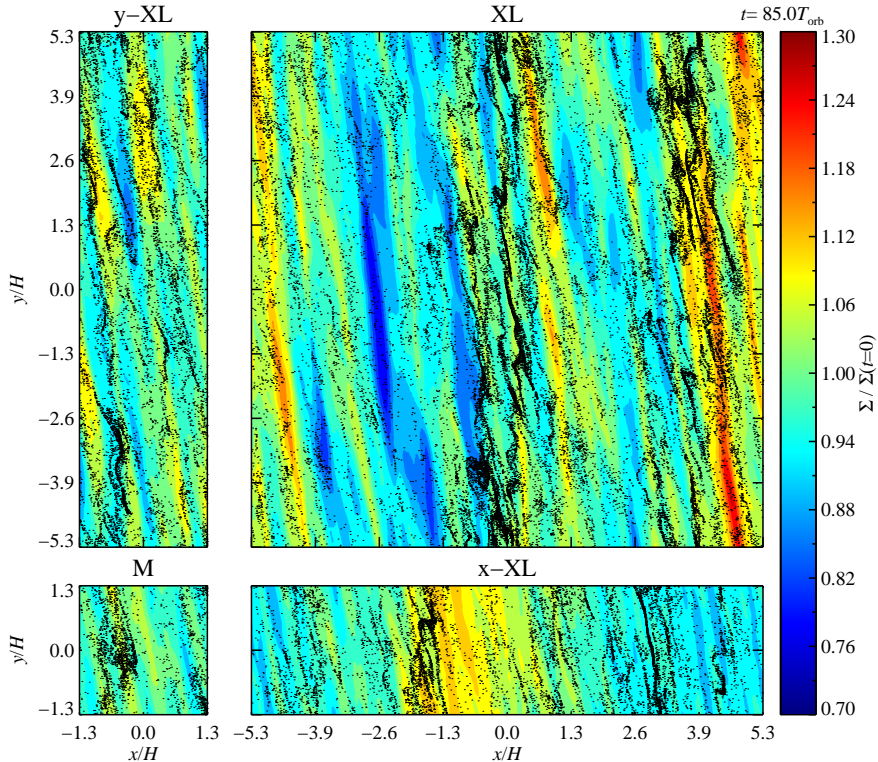}}
  \caption{Collage of four gas surface density representations of the
    \runsss{y-XL}{XL}{M}{x-XL}. Each snapshot was taken after $85 T_{\textrm
      orb}$. These plots show that gas overdensities are most pronounced
    in the largest box. The non-axisymmetric structures have very short
    lifetimes; less than a tenth of an orbit. The pressure bump
    structures are more visible when the density is averaged over the azimuthal
    direction (\Fig{surfcollav}). The black dots represent the position of
    every $100\textrm{th}$ particle, integrated in vertical direction. The
    particles are trapped both in axisymmetric pressure bumps and in spiral
    density waves as described in \citet{HP09}.}
  \label{surfcoll}
\end{figure*}

\begin{figure*}[htb]
  \centering{\includegraphics[width=0.65\linewidth]{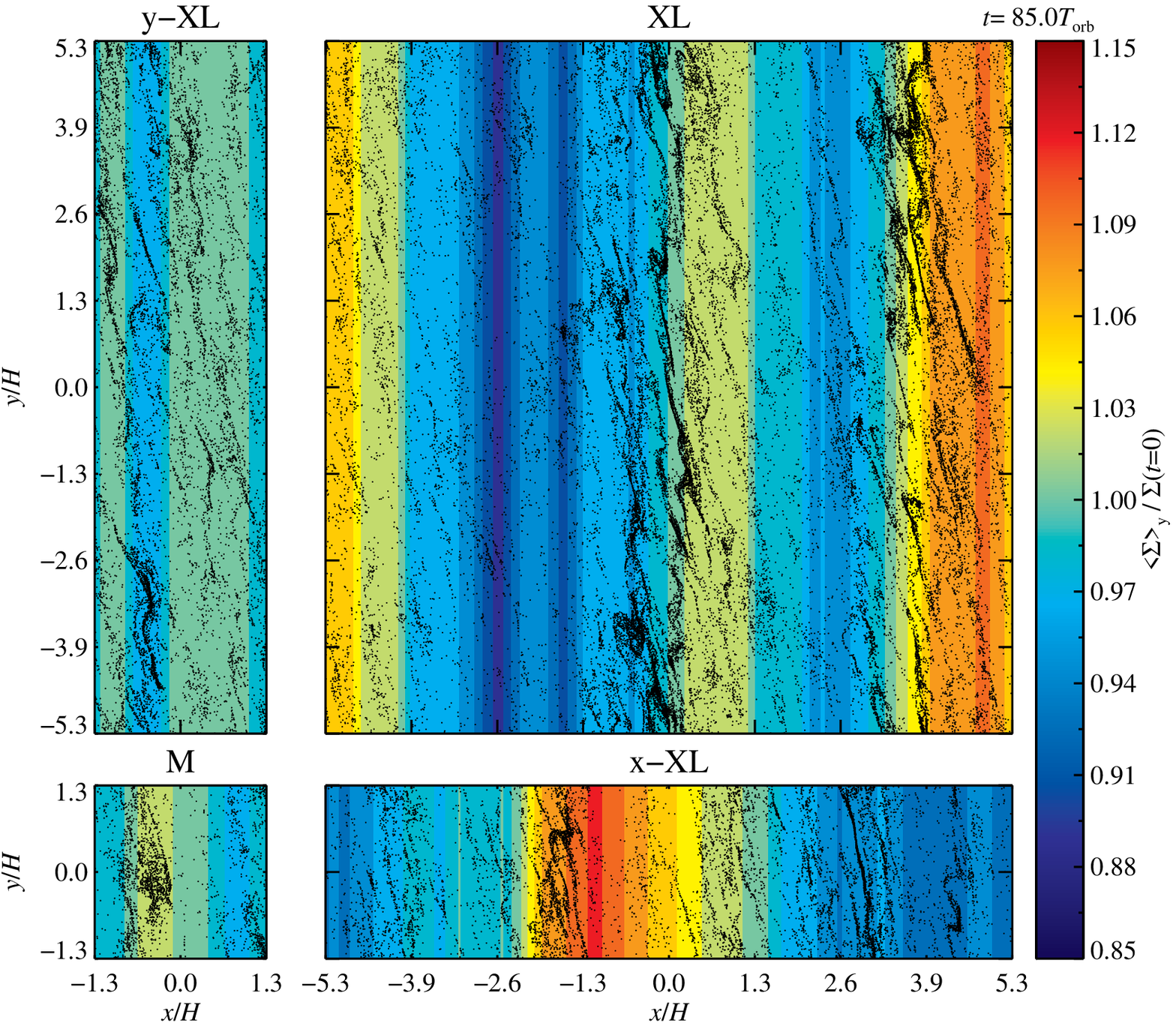}}
  \caption{Surface density distribution of \Fig{surfcoll}, averaged in
    azimuthal direction and averaged over the mean surface density. This
    reveals axisymmetric pressure bumps and valleys. Particles 
    are trapped on the inner side of the density maxima, at places
    with a positive density gradient to overcome the negative global
    pressure gradient. These pressure bumps are stable for many orbits
    (compare \Figs{zonalall}{zonalXXL}).}
  \label{surfcollav}
\end{figure*}

In \Fig{surfcoll} a snapshot of the \runsss{y-XL}{XL}{M}{x-XL} are shown in
scale, giving a real size comparison of high- and low-pressure regions. The
large-scale sinusoidal form of the dominant mode is observable in these
plots. The higher modes are much shorter lived and seem to be non-axisymmetric
density waves affected by the shear \citep{HP09}. The amplitudes of the
pressure differences are higher in azimuthally large boxes. Only the
axisymmetric density waves are long-lived and strong enough to make up a
significant contribution to the pressure bump structure in an azimuthal as
well as a temporal average over some local orbits. The azimuthal average of
\Fig{surfcoll} is seen in \Fig{surfcollav}. Here, the axisymmetric structure
is clearly visible and a strong correlation between the particle location and
a positive radial gradient of the gas density is seen. The black dots in
\Figs{surfcoll}{surfcollav} show the radial and azimuthal position of every
$100\textrm{th}$ particle. The particles are trapped by the axisymmetric
pressure bumps. Also, the shapes of spiral density waves \citep{HP09} can be
seen in the structures. The particle distribution with respect to the gas flow
will be discussed in more detail in \Sec{results2}.

\begin{table*}[htb]
  \centering
    \caption{Zonal Flow Properties} \label{zonalresults}
    \begin{tabular}{lcccccccccc}
      \hline \hline
      Run &
      $\rho_{\textrm{rms}}$ &
      $| \hat{\rho} (k_x = 1)|$ &
      $| \hat{\rho} (k_x = 2)|$ &
      $| \hat{\rho} (k_x = 3)|$ &
      \hspace{0.1cm} & 
      $| \hat{\tilde{u}}_y (k_x = 1)|$ &
      $| \hat{\tilde{u}}_y (k_x = 2)|$ &
      $| \hat{\tilde{u}}_y (k_x = 3)|$ &
      \hspace{0.1cm} & 
      $\tau_{\textrm{corr}}$ \\
      \multicolumn{1}{c}{(1)} & \multicolumn{1}{c}{(2)} &
      \multicolumn{1}{c}{(3)} & \multicolumn{1}{c}{(4)} &
      \multicolumn{1}{c}{(5)} & \hspace{0.1cm} & \multicolumn{1}{c}{(6)} &
      \multicolumn{1}{c}{(7)} & \multicolumn{1}{c}{(8)} & \hspace{0.1cm} &
      \multicolumn{1}{c}{(9)} \\
      \hline
      S &
      $6.1 \times 10^{-3}$ & $4.1 \times 10^{-3}$ &
      $7.5 \times 10^{-4}$ & $4.0 \times 10^{-4}$ &
      \hspace{0.1cm} & $9.9 \times 10^{-3}$ &
      $3.4 \times 10^{-3}$ & $2.1 \times 10^{-3}$ &
      \hspace{0.1cm} & $ 7.6$ \\
      M &
      $2.0 \times 10^{-2}$ & $1.0 \times 10^{-2}$ &
      $2.5 \times 10^{-3}$ & $1.6 \times 10^{-3}$ &
      \hspace{0.1cm} & $1.2 \times 10^{-2}$ &
      $5.2 \times 10^{-3}$ & $3.1 \times 10^{-3}$ &
      \hspace{0.1cm} & $11.2$ \\
      L &
      $3.9 \times 10^{-2}$ & $2.1 \times 10^{-2}$ &
      $5.6 \times 10^{-3}$ & $3.2 \times 10^{-3}$ &
      \hspace{0.1cm} & $1.3 \times 10^{-2}$ &
      $6.5 \times 10^{-3}$ & $3.7 \times 10^{-3}$ &
      \hspace{0.1cm} & $23.2$ \\
      XL &
      $4.3 \times 10^{-2}$ & $1.5 \times 10^{-2}$ &
      $1.1 \times 10^{-2}$ & $5.0 \times 10^{-3}$ &
      \hspace{0.1cm} & $4.6 \times 10^{-3}$ &
      $6.6 \times 10^{-3}$ & $4.3 \times 10^{-3}$ &
      \hspace{0.1cm} & $43.2$ \\
      XXL &
      $4.0 \times 10^{-2}$ & $5.0 \times 10^{-3}$ &
      $7.9 \times 10^{-3}$ & $7.9 \times 10^{-3}$ &
      \hspace{0.1cm} & $7.8 \times 10^{-4}$ &
      $2.4 \times 10^{-3}$ & $3.5 \times 10^{-3}$ &
      \hspace{0.1cm} & $47.3$ \vspace{0.1cm} \\

      x-S &
      $1.6 \times 10^{-2}$ & $3.9 \times 10^{-3}$ &
      $1.7 \times 10^{-3}$ & $1.2 \times 10^{-3}$ &
      \hspace{0.1cm} & $8.0 \times 10^{-3}$ &
      $3.2 \times 10^{-3}$ & $2.0 \times 10^{-3}$ &
      \hspace{0.1cm} & $ 4.4$ \\
      x-L &
      $3.3 \times 10^{-2}$ & $2.3 \times 10^{-2}$ &
      $7.3 \times 10^{-3}$ & $2.7 \times 10^{-3}$ &
      \hspace{0.1cm} & $1.4 \times 10^{-2}$ &
      $8.8 \times 10^{-3}$ & $4.7 \times 10^{-3}$ &
      \hspace{0.1cm} & $37.6$ \\
      x-XL &
      $2.7 \times 10^{-2}$ & $1.2 \times 10^{-2}$ &
      $1.0 \times 10^{-2}$ & $6.4 \times 10^{-3}$ &
      \hspace{0.1cm} & $3.5 \times 10^{-3}$ &
      $6.2 \times 10^{-3}$ & $5.6 \times 10^{-3}$ &
      \hspace{0.1cm} & $20.2$ \vspace{0.1cm} \\

      y-S &
      $1.2 \times 10^{-2}$ & $9.9 \times 10^{-3}$ &
      $2.4 \times 10^{-3}$ & $1.1 \times 10^{-3}$ &
      \hspace{0.1cm} & $1.2 \times 10^{-2}$ &
      $5.9 \times 10^{-3}$ & $4.1 \times 10^{-3}$ &
      \hspace{0.1cm} & $14.4$ \\
      y-L &
      $3.0 \times 10^{-2}$ & $6.9 \times 10^{-3}$ &
      $3.8 \times 10^{-3}$ & $3.1 \times 10^{-3}$ &
      \hspace{0.1cm} & $7.8 \times 10^{-3}$ &
      $3.7 \times 10^{-3}$ & $2.3 \times 10^{-3}$ &
      \hspace{0.1cm} & $10.8$ \\
      y-XL &
      $3.6 \times 10^{-2}$ & $5.6 \times 10^{-3}$ &
      $4.1 \times 10^{-3}$ & $3.3 \times 10^{-3}$ &
      \hspace{0.1cm} & $5.7 \times 10^{-3}$ &
      $2.6 \times 10^{-3}$ & $1.6 \times 10^{-3}$ &
      \hspace{0.1cm} & $10.3$ \vspace{0.1cm} \\

      LspecMR &
      $3.7 \times 10^{-2}$ & $1.8 \times 10^{-2}$ &
      $5.5 \times 10^{-3}$ & $3.1 \times 10^{-3}$ &
      \hspace{0.1cm} & $1.1 \times 10^{-2}$ &
      $6.1 \times 10^{-3}$ & $3.6 \times 10^{-3}$ &
      \hspace{0.1cm} & $21.8$ \\
      LspecHR &
      $4.2 \times 10^{-2}$ & $9.8 \times 10^{-3}$ &
      $3.2 \times 10^{-3}$ & $2.0 \times 10^{-3}$ &
      \hspace{0.1cm} & $5.8 \times 10^{-3}$ &
      $3.6 \times 10^{-3}$ & $2.0 \times 10^{-3}$ &
      \hspace{0.1cm} & $23.4$ \\
      LspecMRs &
      $4.3 \times 10^{-2}$ & $2.4 \times 10^{-2}$ &
      $5.4 \times 10^{-3}$ & $3.2 \times 10^{-3}$ &
      \hspace{0.1cm} & $1.4 \times 10^{-2}$ &
      $6.1 \times 10^{-3}$ & $3.9 \times 10^{-3}$ &
      \hspace{0.1cm} & $10.9$ \\
      MspecMRb &
      $1.9 \times 10^{-2}$ & $8.6 \times 10^{-3}$ &
      $2.6 \times 10^{-3}$ & $1.6 \times 10^{-3}$ &
      \hspace{0.1cm} & $1.0 \times 10^{-2}$ &
      $5.3 \times 10^{-3}$ & $3.1 \times 10^{-3}$ &
      \hspace{0.1cm} & $26.4$ \vspace{0.1cm} \\

      L\_SAFI &
      $3.9 \times 10^{-2}$ & $2.1 \times 10^{-2}$ &
      $5.3 \times 10^{-3}$ & $3.2 \times 10^{-3}$ &
      \hspace{0.1cm} & $1.2 \times 10^{-2}$ &
      $6.1 \times 10^{-3}$ & $3.9 \times 10^{-3}$ &
      \hspace{0.1cm} & $25.6$ \\
      XL\_SAFI &
      $4.8 \times 10^{-2}$ & $2.4 \times 10^{-2}$ &
      $1.0 \times 10^{-2}$ & $5.4 \times 10^{-3}$ &
      \hspace{0.1cm} & $7.0 \times 10^{-3}$ &
      $6.2 \times 10^{-3}$ & $4.8 \times 10^{-3}$ &
      \hspace{0.1cm} & $48.6$ \\
      \hline
    \end{tabular}
  \\{\textbf Notes.} Column 1: name of run. Column 2: root-mean-square
  density $\rho_{\textrm{rms}} = \sqrt{ \langle ( \rho - \overline{\rho} )^2
    \rangle}$. Columns 3-5: Fourier amplitude of radial density modes $k_x = 1
  \ldots 3$, normalized by mean density in the box. Columns 6-8: Fourier
  amplitude of azimuthal velocity modes $k_x = 1 \ldots 3$ with $\tilde{u}_y =
  u_y - \overline{u}_y$. Column 9: correlation time, in orbits $T = 2 \pi
  \Omega^{-1}$, of the largest radial density mode. 
\end{table*}

\begin{figure*}[t!]
  \centering
    \includegraphics[width=\linewidth]{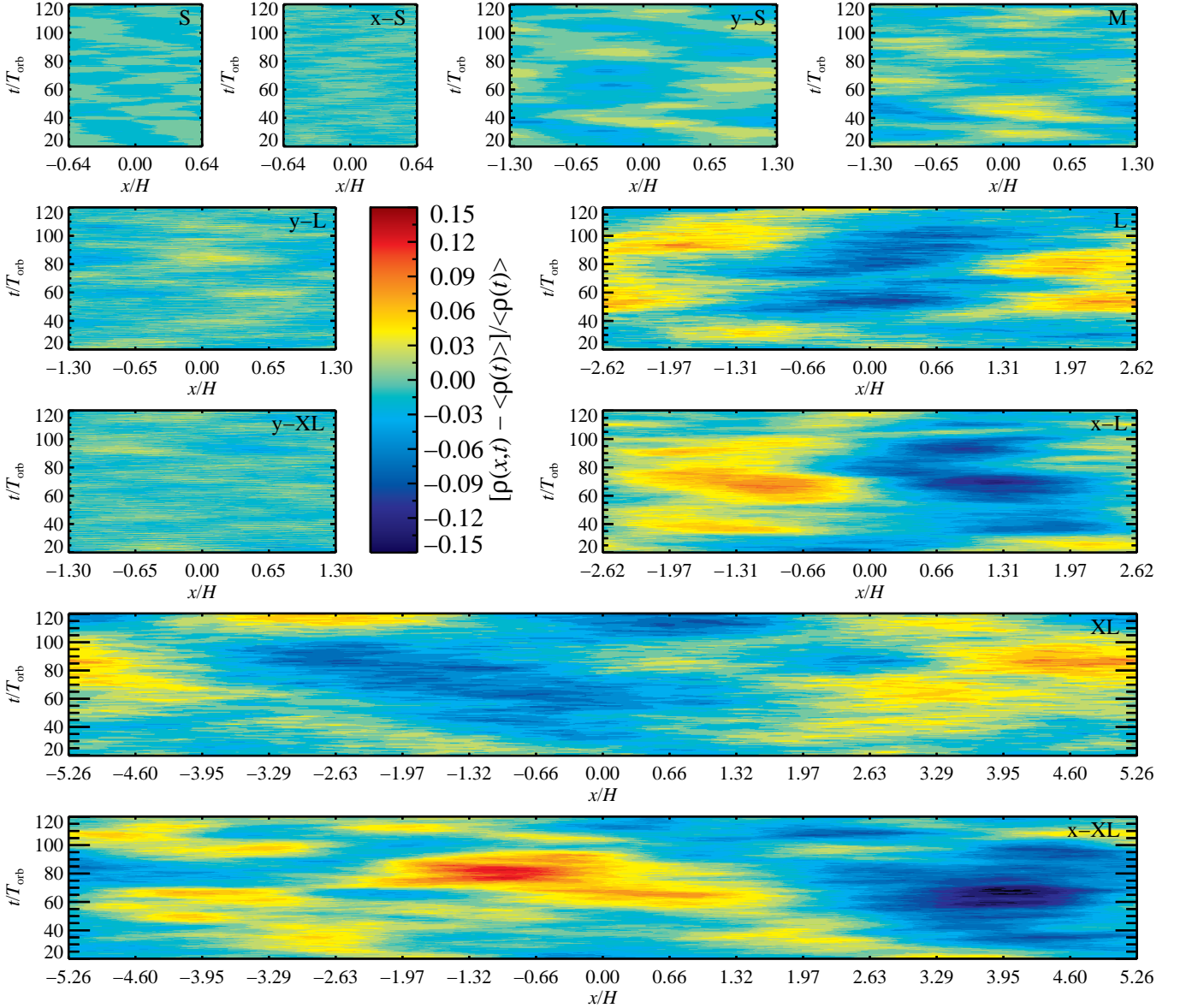}
  \caption{Evolution of the gas density perturbation of all runs from
    simulation sets \textit{A}, \textit{B}, and \textit{C}. Run~\textit{XXL}
    is shown in \Fig{zonalXXL}. The density is averaged in vertical and
    azimuthal direction and plotted in radial direction over time. The
    lifetime, the size, as well as the strength of the pressure bumps are
    clearly increasing with increasing box size in simulation set \textit{A},
    i.e., \runssss{S}{M}{L}{XL}{XXL}. In simulation set \textit{B}, i.e.,
    \runsss{x-S}{M}{x-L}{x-XL}, we have the same increase of lifetime, size,
    and strength of the pressure bumps. Only for the very large simulation
    ($L_x=10.56H$), there is no apparent difference in pressure bump size and
    strength to \run{x-L}. For simulation set \textit{C}, i.e.,
    \runsss{y-S}{M}{y-L}{y-XL}, the strength  of the pressure bumps is
    apparently constant throughout this set of simulations. Even the lifetime
    decreases slightly with increasing box size.}
  \label{zonalall}
\end{figure*}

\begin{figure*}[t!]
  \centering
    \includegraphics[width=0.49\linewidth]{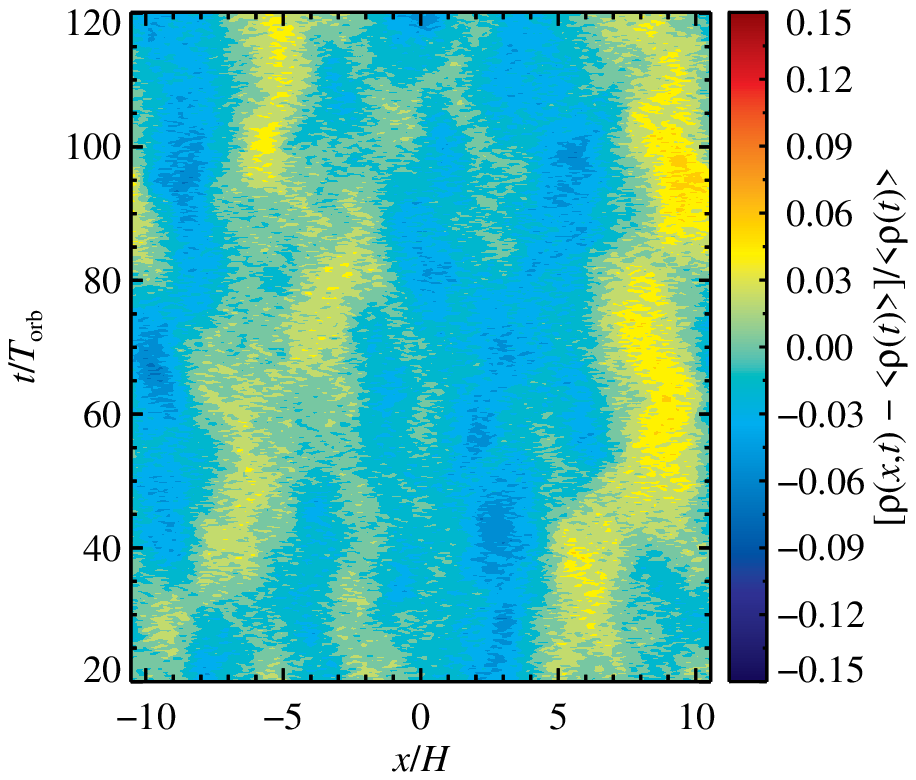}
    \includegraphics[width=0.49\linewidth]{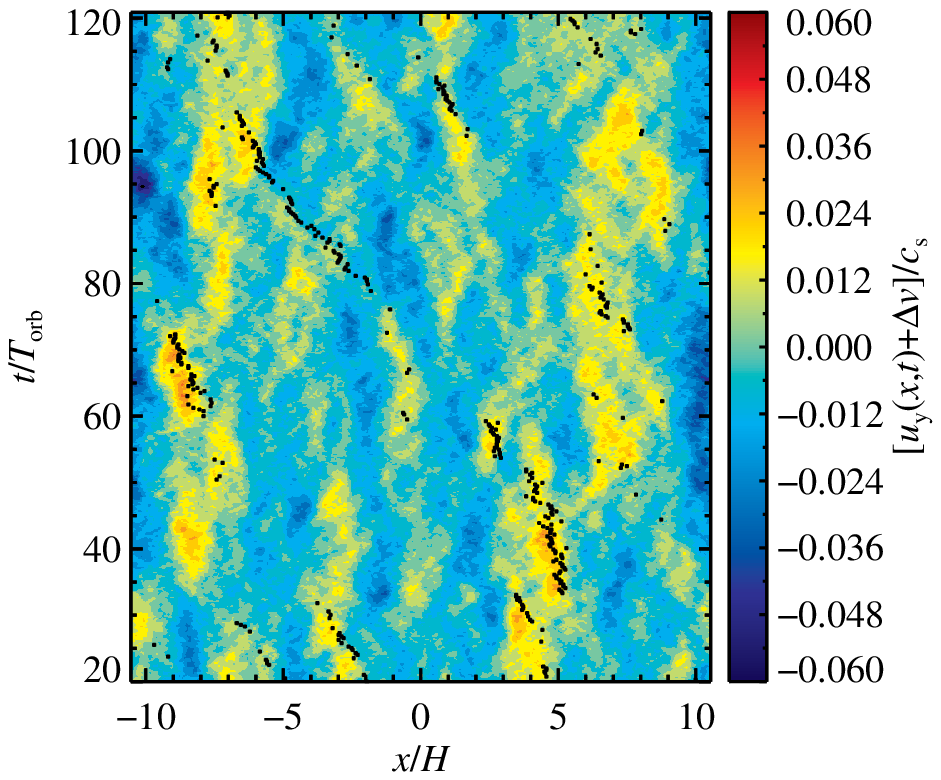}
    \includegraphics[width=0.49\linewidth]{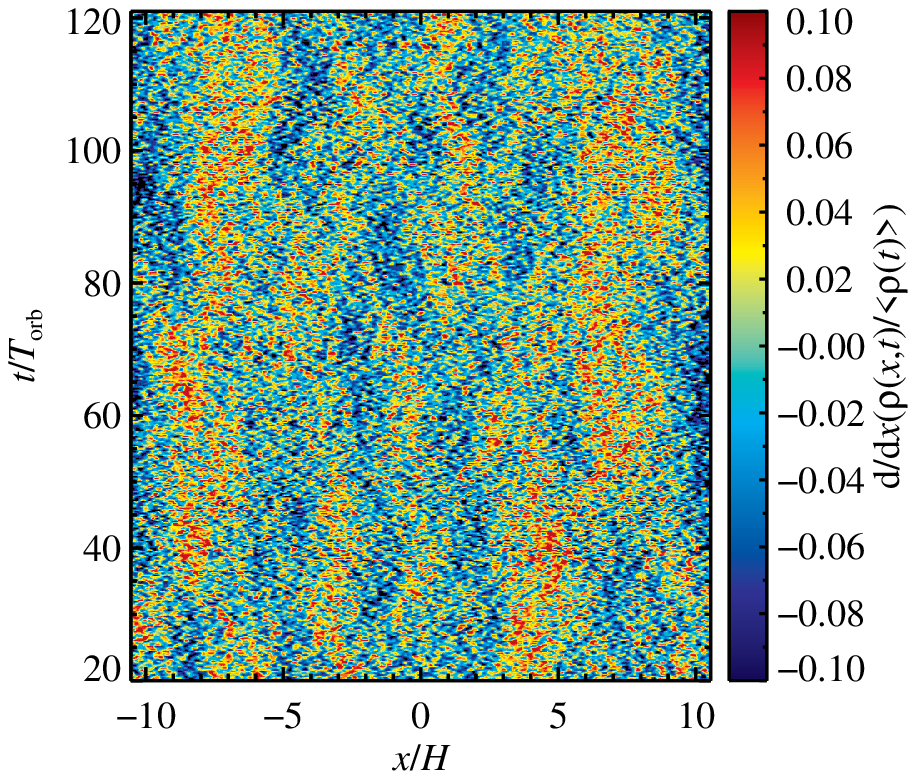}
    \includegraphics[width=0.49\linewidth]{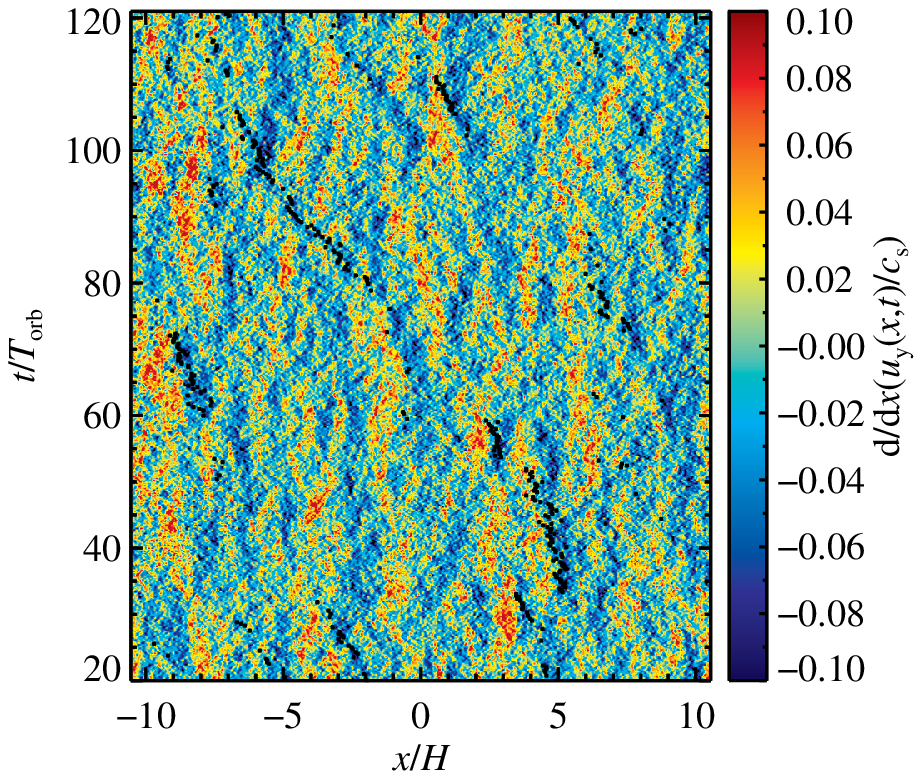}
  \caption{Top panels show the evolution of the gas density perturbation
    and the azimuthal gas velocity of \run{XXL}, whereas the bottom row	shows
    the radial derivative of these quantities. The derivatives are very
    speckled, since small-scale fluctuations give stronger amplitudes to the
    derivatives. However, the underlying large-scale structure is still
    visible. The azimuthal gas velocity follows the radial gas density
    gradient, as expected for a geostrophic balance. Hence, it is possible to
    interpret the radial derivative of the azimuthal gas velocity as the
    second derivative of the gas density. In the upper right panel, the black
    dots represent the position of the most massive particle clump in the
    simulation at each time. It is clearly shown that particles get trapped in
    regions of positive zonal flow downstream of pressure bumps.}
  \label{zonalXXL}
\end{figure*}

All of our simulations show signs of zonal flows. Strength and lifetime of
the zonal flows and the associated pressure bumps differ very much with the
physical box size. Space--time plots of all different simulation sizes are
shown in \Fig{zonalall} and the upper left panel of \Fig{zonalXXL}. The
pressure bumps are generally more pronounced in simulations with a larger
radial extent. Simulation set \textit{A} strictly follows this general
trend. Pressure bump features grow in strength and lifetime with the physical
box size, always staying at the largest radial scale. This rule applies to all
but the largest \runss{XL}{x-XL}{XXL}. There, instead of the formerly
predominant $k_x = 1$ ($\omega_x = 2 \pi k_x / L_x$ in Fourier mode
$\sin{(\omega_x x)}$) mode, the mode $k_x = 2$ (higher modes for \run{XXL}) is
occupied by the pressure bumps. In simulation set \textit{B}, the strength and
size of the pressure bumps converges for simulations with a radial extent
larger than $5.28H$. The lifetime of the pressure bumps even decreases for
the largest simulation in this set. Simulation set \textit{C} is qualitatively
different from the other simulation sets. Strength, size, and lifetime of the
pressure bumps seem to be inversely proportional to the azimuthal extent of
the box, when the vertical and radial box sizes are kept constant. This effect
was already seen in, e.g., \citet{SBA12} and \citet{FDKTH12}. Both groups show
that the magnetic field consists of two components: a local turbulent
component that is responsible for the zonal flows and a global azimuthal
component. Since the total energy stays approximately constant, the local
component gets weaker and consequently zonal flows as well as axisymmetric
pressure bumps get weaker too.

\Fig{zonalXXL} additionally shows space--time plots of the azimuthal gas
velocity and the radial gradients of gas density and azimuthal gas
velocity of \run{XXL}. In the right panels, the position of the highest
dust density are shown as dots. Particles get clearly slowed by the maxima of
the azimuthal gas velocity, i.e., the large-scale maxima of the pressure
gradients. The velocity has large-scale structures that are very similar to
those of the density gradient, as expected in geostrophic balance. Thus, the
structure of the velocity gradient can be approximated as the large-scale
structure of the second derivative of the gas density. It is shown in the
lower right panel that the particles get stopped at the minima of the radial
derivative of the azimuthal gas velocity (and thus at minima of the second
derivative of the gas density) as analytically predicted \citep[see,
e.g.,][]{KL01}.

We calculated the correlation time of the pressure bumps and the zonal
flows in the same way as it was calculated in \citet{JYK09}. We use the
density $\rho$, averaged over azimuthal and vertical directions, at a given
time $t$. Then, we average over each point in radial direction the time it
takes for the density at each point to change by a value corresponding to
the standard deviation of the gas density. These measurements are taken for
every local orbit. The measurements are averaged over the time between
saturation of the turbulence and a time when the correlation does not
extend the correlation time to the final time of the simulation. Finally,
the averages are multiplied by two, in order to cover the full temporal
extent of the correlated structures. The correlation times measured in this
fashion are in good agreement with the lifetime of the overdensities that
is seen in \Figs{zonalall}{zonalXXL}. However, a change of position of
the structures, as seen in \run{XL} (check \Fig{zonalall}), is not accounted
for. Thus, correlation times are more likely to be underestimated than
overestimated. Also, we cannot be entirely sure whether this behavior is 
really drift or structure decay and reformation.

\begin{figure}[htb]
  \centering
    \includegraphics[width=\linewidth]{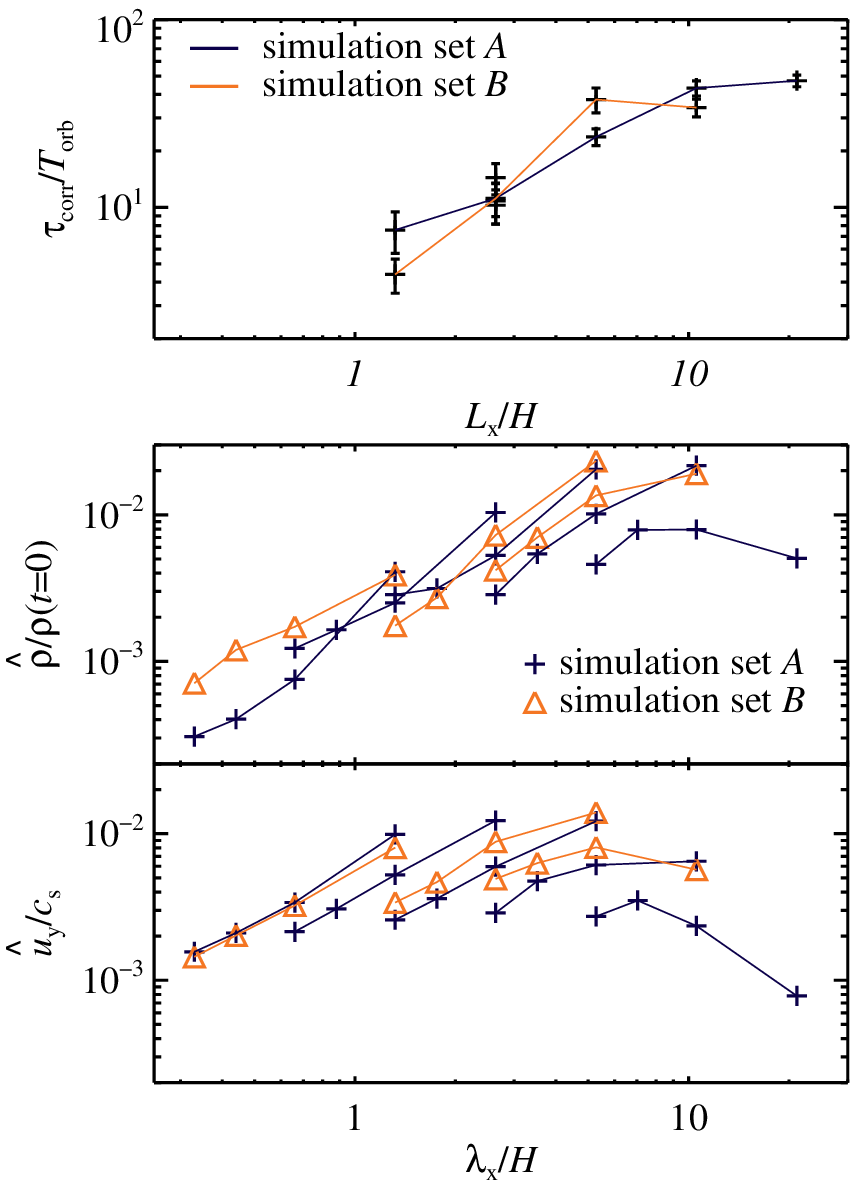}
  \caption{Upper plot shows the correlation times of the largest radial
    density mode against the radial box size. The lines correspond to the
    simulation sets \textit{A} and \textit{B}. The results from simulation set
    \textit{C} are omitted for visibility. The correlation time 
    $\tau_{\text{corr}}$ grows for boxes with a larger radial extent. Only
    \run{x-XL} does not follow this trend. The large ratio $L_x/L_y$ may
    prohibit formation of stable zonal flows. The two lower plots show the
    first four amplitudes of the radial Fourier modes of the gas density and
    the azimuthal gas velocity against their real size $\lambda_x = L_x /
    k_x$; $k_x$ is the wave number of the corresponding Fourier mode, defined
    by $\omega_x = 2 \pi k_x / L_x$ in the Fourier mode $\sin{(\omega_x
      x)}$. The lines connect the amplitudes of different Fourier modes for
    one simulation. Both quantities have most of their power in the largest
    modes. Only in the largest simulations, the power in the largest modes
    decreases. There the maximum is between $5H$ and $7H$.}
  \label{correlationtimes}
\end{figure}

The results of the correlation time determination are shown in
\Tab{zonalresults} and in the upper panel of \Fig{correlationtimes}. For the
diagonal simulation set, (\textit{A}), the correlation time increases with box
size. It seems to saturate toward the largest box size. The trend to longer
correlation times is also evident for simulation set \textit{B}. Here only
\run{x-XL} has a shorter correlation time than expected. This might be an
effect of the strongly stretched simulation box. The correlation time
decreases slightly with an increasing azimuthal box size in simulation set
\textit{C} (not shown in the figure). The lower panel in
\Fig{correlationtimes} shows a measurement for the physical size of the zonal
flow features. We Fourier-transformed the vertically and azimuthally averaged
gas density and azimuthal gas velocity for each time step and averaged the
amplitudes of the first four modes over the time of $20 \ldots 120$ local
orbits. The length was normalized for the size of the simulation box, to get
the physical size of the modes by $\lambda_x = L_x / k_x$ with the wave number
$k_x$. The turbulence is always strongest at the largest modes for simulations
with $L_x \lesssim 5H$. The highest amplitude for both quantities in the
largest simulation domain is found between $5H$ and $7H$ (up to $10H$ for
$\hat{\rho}$). These measurements are also found in \Tab{zonalresults}.

The \runs{L\_SAFI}{XL\_SAFI} were carried out to compare the turbulence and
zonal flow parameters with the \runs{L}{XL}. They were run to check that zonal
flows are no effects from the shear advection scheme that was used in the Pencil
Code. Comparing the values in \Tabs{turbresults}{zonalresults} shows that
there is little change in the measured properties of the zonal flows and the
associated pressure bumps. However, the computation time increases if one uses
the SAFI scheme. Thus, this scheme was only used to confirm our results.

\section{Particle Behavior in Zonal Flows}\label{results2}

Particle accumulations and planetesimal formation can occur in clumps and
filaments of the overdensities in the dust. In our simulations, we do not
include gravitational interaction between the particles. Thus, we only
study the passively developed overdensities of the dust to see when and
whether overdensities sufficient for the streaming instability can be
reached. By not having explicit feedback one can retroactively study the
concentration factor for various initial dust-to-gas ratios. Simulations
including feedback will have to be done in future studies. \Fig{surfcoll}
shows the position of every $100\textrm{th}$ particle in selected
simulations. These plots clearly show the trend for particles to accumulate in
the downstream of high-pressure regions. Particles are pulled toward pressure
gradient maxima \citep{KL01}. In the upper right panel in \Fig{surfcoll}, a
snapshot of \run{XL} after $85$ local orbits is shown. The particles clump up
at positions just left of the maxima in the $k_x = 1$ of the gas density;
these are the locations of positive zonal flows, i.e., regions where the
azimuthal gas velocity is higher than the pressure-supported Keplerian flow.

\subsection{Particles in Zonal Flows}\label{zonalpart}

In the upper right panel of \Fig{zonalXXL}, the azimuthal gas velocity
development of \run{XXL} is shown, overplotted with the position of the most
massive clump for each time step. The azimuthal gas velocity coincides with
the derivative of the gas density, but it is much easier to interpret. The
speckled structure of the derivatives comes due to the high power in the
smaller scales. However, the large-scale structure is still visible and the
geostrophic correlation between the structures of $u_y(x,t)$ and
$d/dx[\rho(x,t)]$ is directly observed. Since they have the same
large-scale structure, the particle position is much easier interpreted at
the azimuthal gas velocity plot than on the density gradient plot. Sometimes
the radial displacement from one orbit to the next is too large to be
explained by radial drift. That happens when another clump becomes more
massive than the previous one. These particles accumulate in regions with high
azimuthal gas velocities (see upper right panel in \Fig{zonalXXL}). The only
time when this is not true is at times from $80$ to $100$ local orbits. In
this period, an inward-drifting clump stayed coherent during the time of its
drift. The drift velocity of the most massive particle clump is indirectly
encrypted in this plot. Particles are drifting much slower when they are
trapped by a pressure gradient. As all particles drift inward this leads to
accumulation of particles in regions where the perturbed pressure gradient is
positive.

\begin{figure}[htb]
  \centering
    \includegraphics[width=\linewidth]{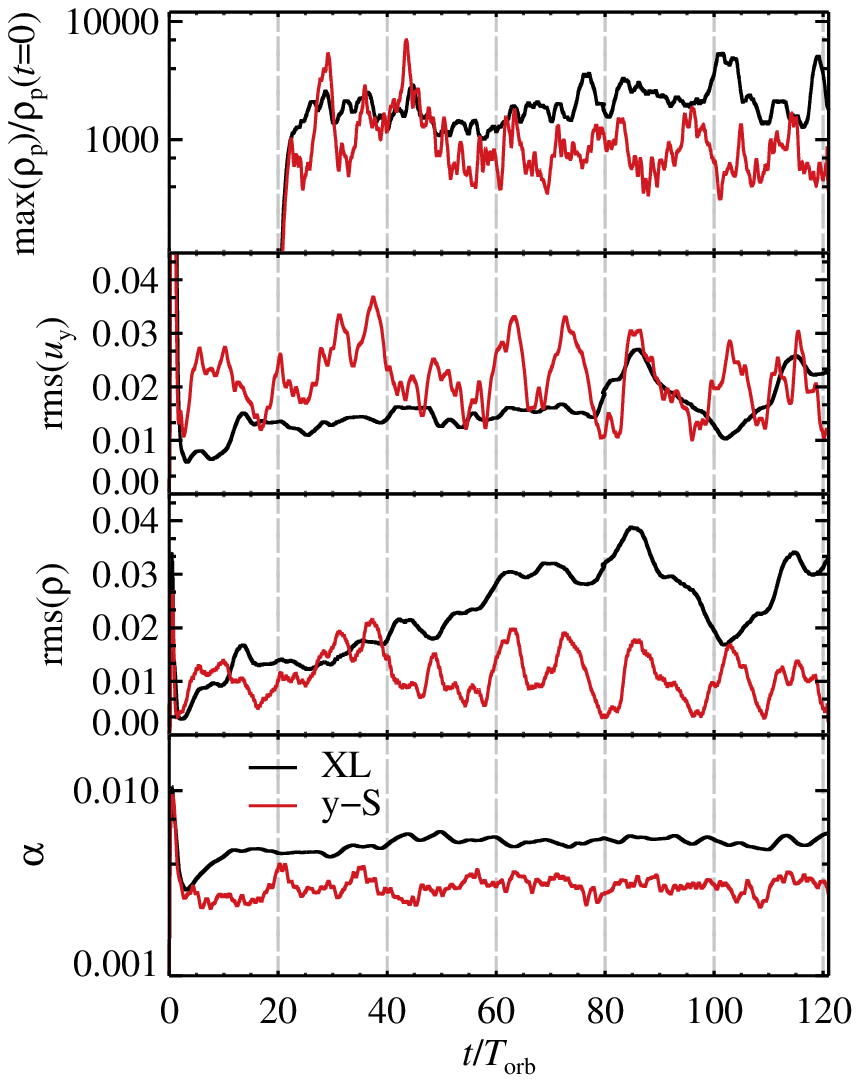}
  \caption{Time series of \runs{XL}{y-S}. The plots show (from top to bottom)
    the maximum of the dust density, the root mean square of the azimuthal gas
    velocity and the gas density, and the $\alpha$-value (\Eq{eqnalpha}). The
    dust overdensities of \run{XL} have a higher base than those of
    \run{y-S}. The latter has some spikes in the beginning, but is lower for
    most time of the simulation. The two panels in the middle show that the
    azimuthal gas velocity and the gas density are correlated. Both plots show
    maxima and minima at the same time, while $\alpha$ is rather stable with
    time. The time-averaged $\alpha$-values for all simulations can be found
    in \Tab{turbresults}.}
  \label{timeseries}
\end{figure}

The maximal accumulation of particles for \runs{XL}{y-S} are plotted in the
top panel in \Fig{timeseries}. The second panel shows the evolution of the
quantity $\sqrt{\langle u_y - \langle u_y \rangle_{yz} \rangle_x^2}$, a
measure for the strength of the zonal flows. The third panel in
\Fig{timeseries} shows the evolution of the strength of the gas density
enhancement as $\sqrt{\langle \rho - \langle \rho \rangle_{yz}
  \rangle_x^2}$. Comparing the second and third panels, one can see a clear
correlation between the zonal flow strength and the gas density
enhancements. The bottom panel in \Fig{timeseries} shows the evolution of the
$\alpha$-parameter, calculated as in \Eq{eqnalpha}.

\begin{figure}[htb]
  \centering
    \includegraphics[width=\linewidth]{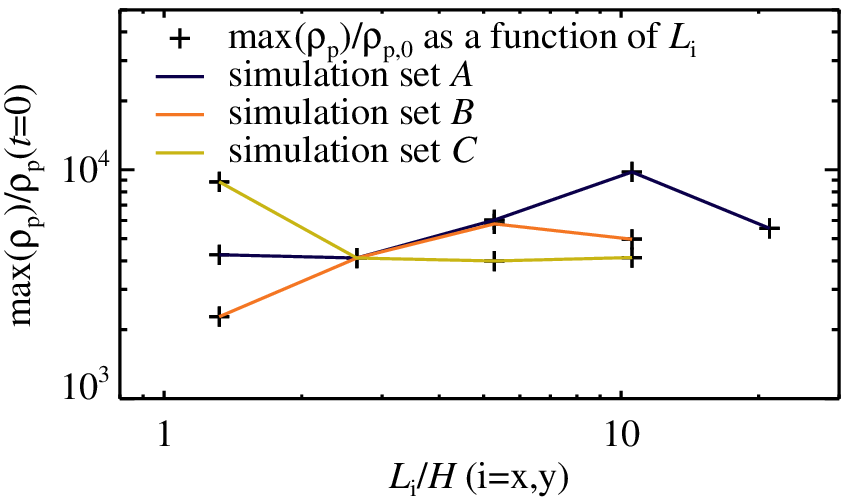}
    \includegraphics[width=\linewidth]{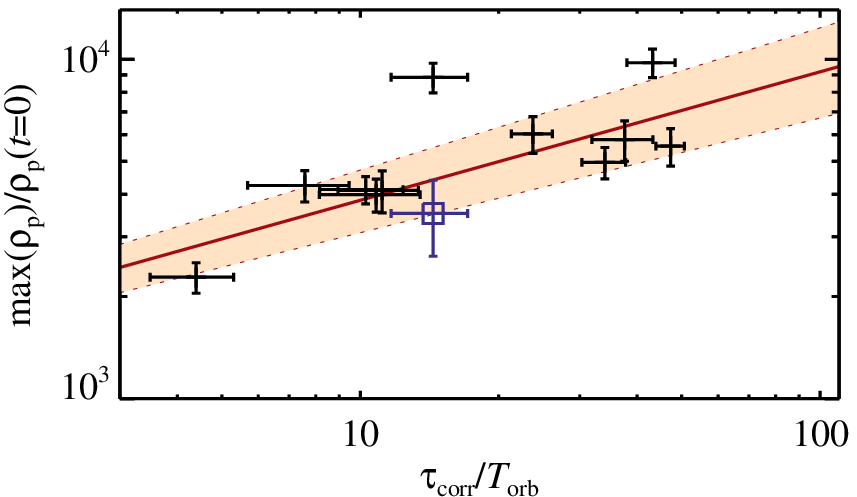}
  \caption{In the upper panel, the highest peak in the time series of the
    dust overdensity (the top panel in \Fig{timeseries}) is plotted against
    the size of the simulation box. The diagonal simulation set \textit{A} is
    marked by the blue line. The dust overdensity increases with box size,
    until it suddenly drops for the largest box. In simulation set \textit{B}
    (red), the quantity saturates for boxes with a radial extent that is
    twice as large as the azimuthal extent or larger. When keeping the radial
    extent constant (simulation set \textit{C}, yellow line), the maximum
    saturates for the cubic box case. Hence, the only in azimuthal direction
    extended boxes do not lead to an artificial enhancement of the dust
    overdensities. The very high overdensity for \run{y-S} seems to be a
    stochastic coincidence (compare top panel in \Fig{timeseries}). The
    lower panel shows the dust overdensity against the correlation time.
    The measured points can be approximated with a power law (shown as a red
    solid line) with an exponent of $0.38 \pm 0.05$. The shaded region, with
    the dashed red lines as edges, gives the uncertainty of the fit. The one
    cross off the fit shows the results for \run{y-S}. The blue square marks
    its position, if we neglect the two maxima shown in \Fig{timeseries}. It
    overlaps well with the fit region.}
  \label{rhopmax}
\end{figure}

The maximum of the dust overdensity that occurs during one simulation is
plotted against the box size in the upper panel of \Fig{rhopmax}. The
general trend shows that radially larger boxes have higher particle
concentrations. An increased azimuthal extent does not have an effect on the
particle concentrations. The most surprising result is in \run{y-S}. It
shows a very high particle concentration that occurs early in the simulation
(compare \Fig{timeseries}). This is most likely a stochastic coincidence.
The lower panel in \Fig{rhopmax} shows a plot of the maximum dust
overdensity against the correlation time of the zonal flows. The error margin
are calculated with the standard deviation of the temporal evolution of the
two quantities. We see a clear trend that denser particle accumulations
develop with longer correlation times. The distribution can be fitted by a
power law. This gives an exponent of $d\log{\rho}/d\log{\tau_{\textrm{corr}}}
= 0.38 \pm 0.05$. The one point that does not overlap with the error margins
of the fit is from \run{y-S}. If we take the maximum of the top panel in
\Fig{timeseries} after the two first maxima (i.e., after $45 \Torb$) and plot
this value again in the parameter space of \Fig{rhopmax}, we get the position
marked with the blue square. It agrees well with the error margins of the
fit.

\begin{figure*}[t!]
  \centering
    \includegraphics[width=0.49\linewidth]{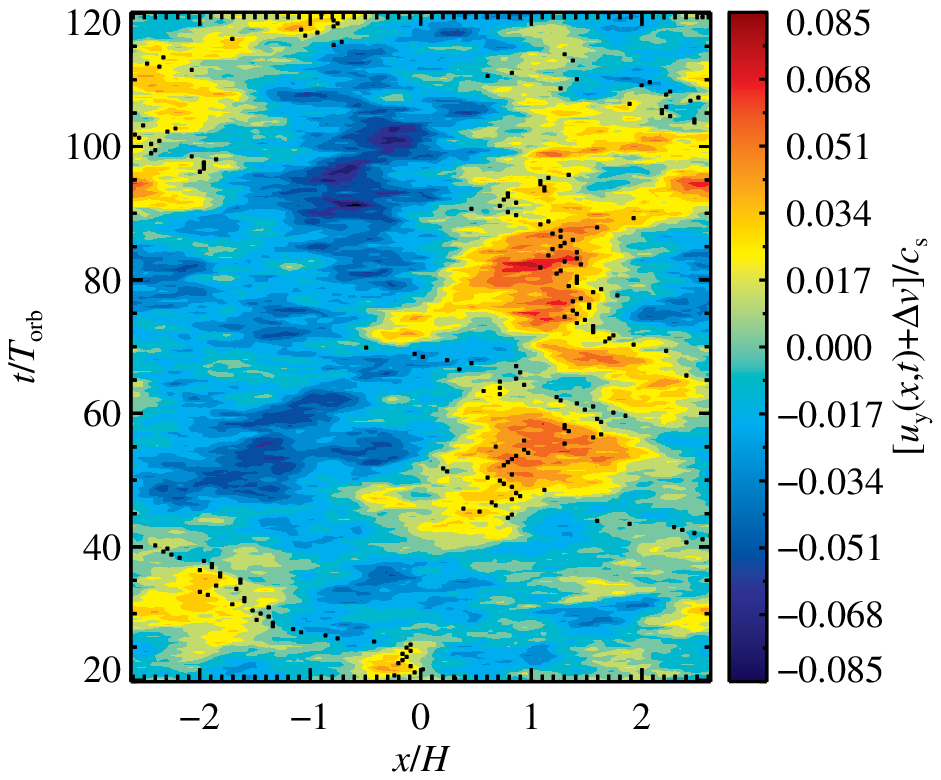}
    \includegraphics[width=0.49\linewidth]{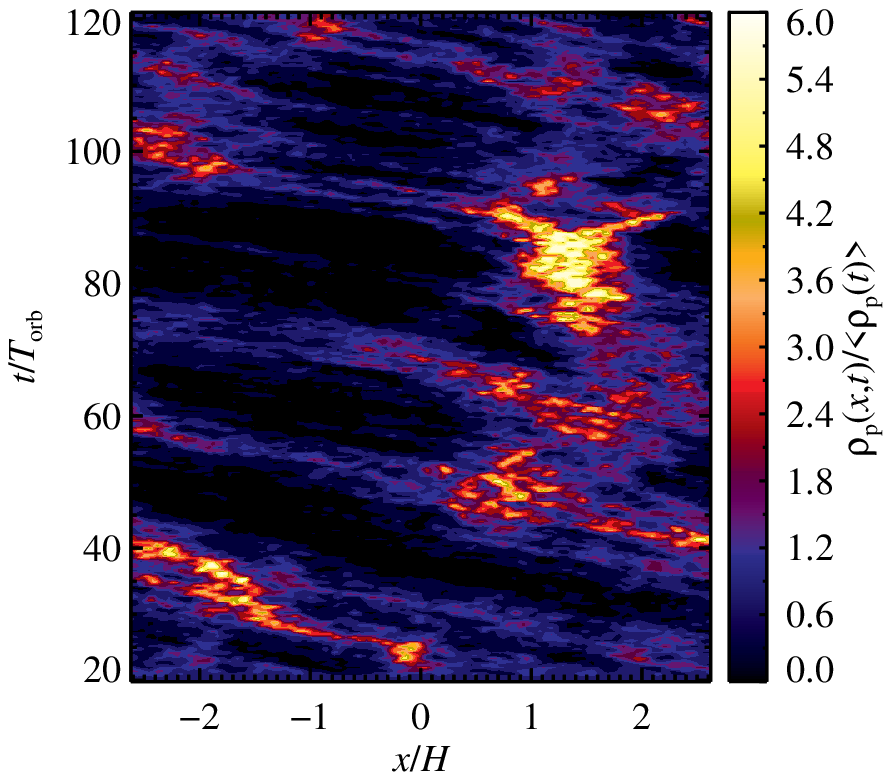}
    \includegraphics[width=0.49\linewidth]{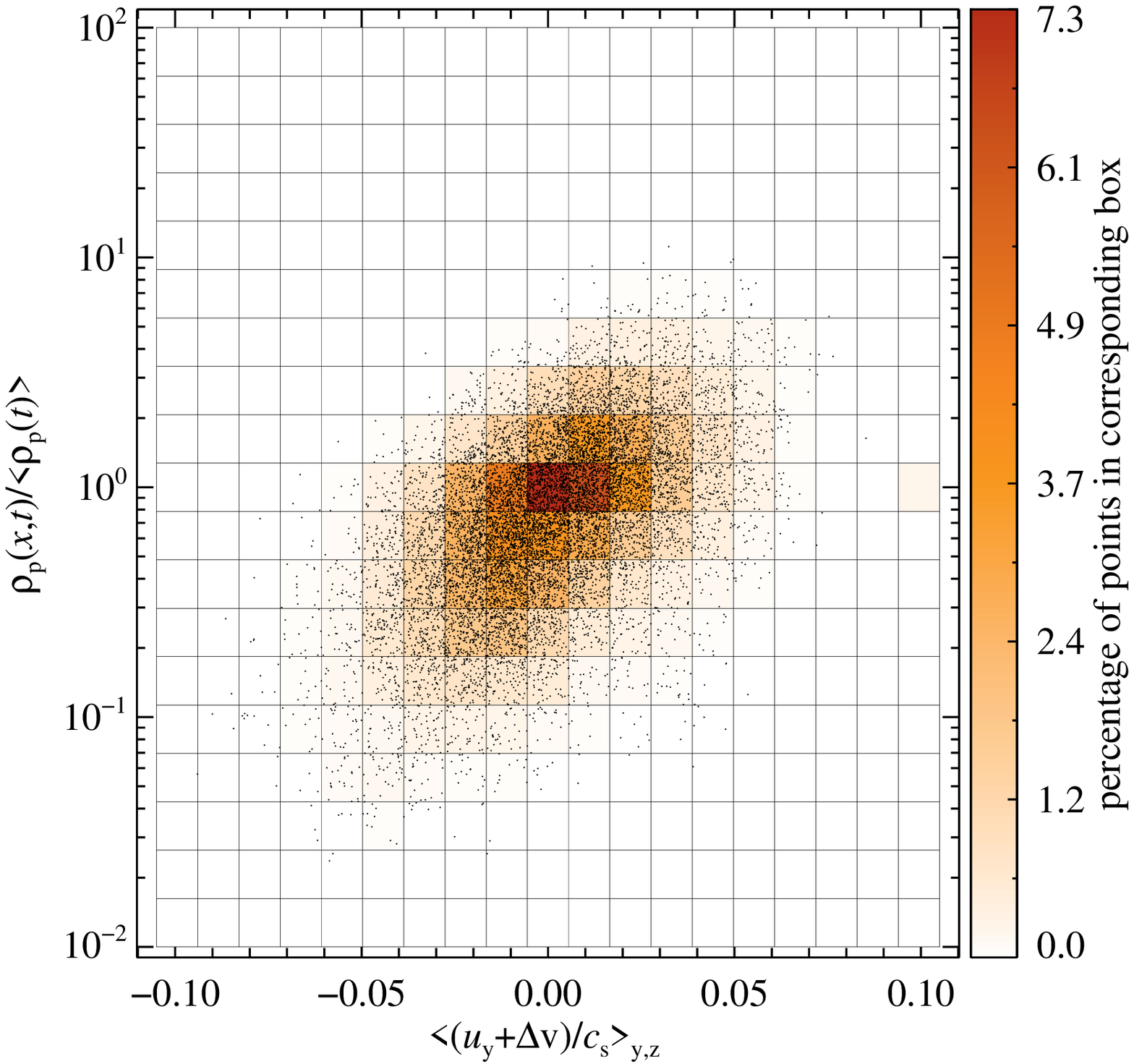}
    \includegraphics[width=0.49\linewidth]{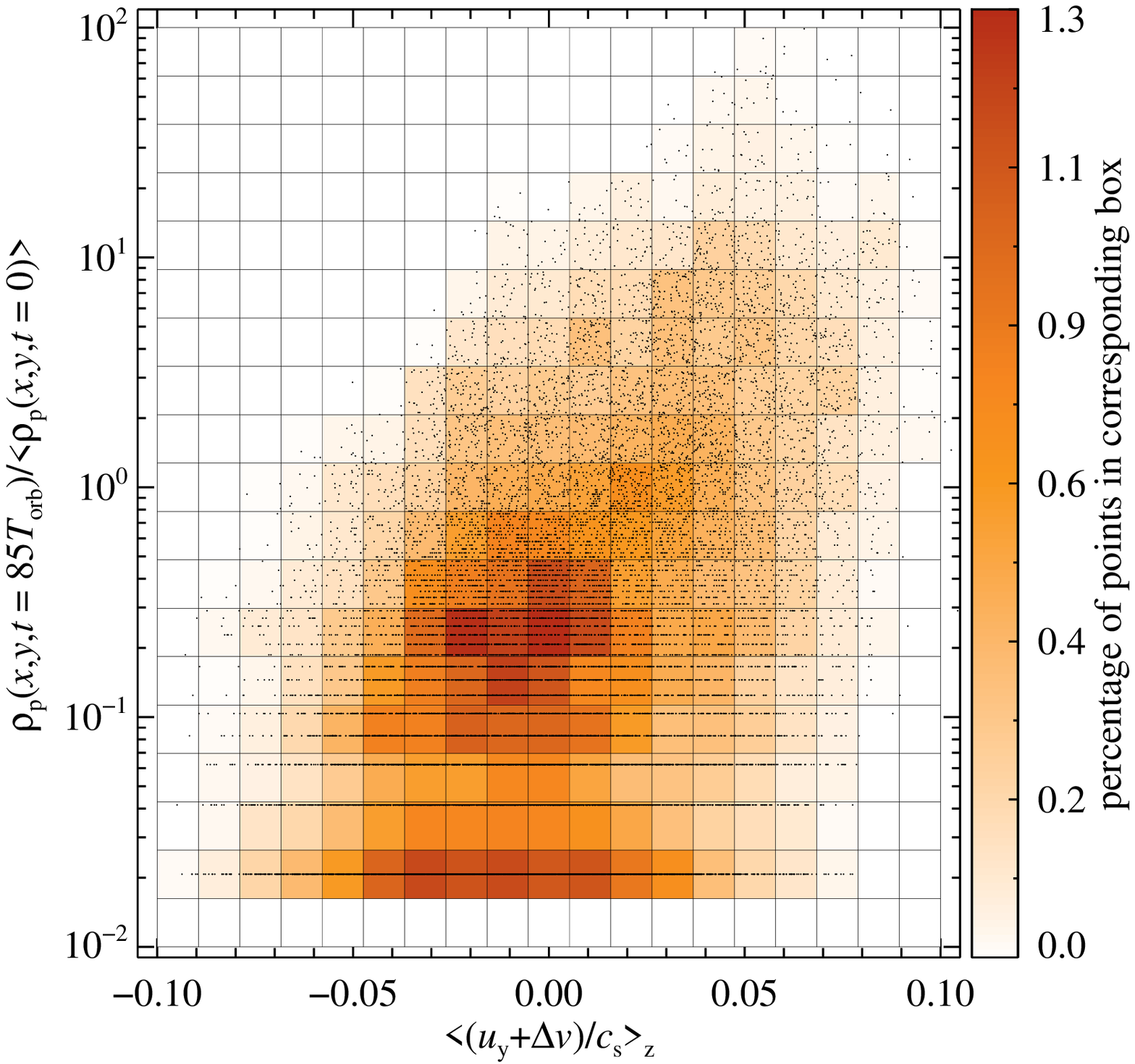}
  \caption{Top row shows the evolution of the azimuthal gas velocity and
    the dust density evolution of \run{L}, respectively. The quantities are
    averaged in vertical and azimuthal direction and plotted in radial
    direction over time. The black dots in the upper left panel show the
    position of the highest dust density at each orbit. This shows that the
    overdensities of the dust often appear at places and times where the
    azimuthal gas velocity is high. This relation shows that the zonal flows
    accumulate dust and are a possible venue of planetesimal formation. The
    bottom left panel shows a scatter plot of the dust density
    $\rho_{\textrm{p}}$ against the azimuthal gas velocity, where both values
    are averaged in vertical and azimuthal direction, as in the upper
    panels. The bottom right panel shows a plot of the particle surface
    density $\rho_{\textrm{p}}(x,y,t)$ in relation to the azimuthal gas
    velocity, averaged in vertical  direction, computed from a snapshot taken
    at $85\Torb$, the time when the maximum in the dust density occurs. Both
    plots show that it is more likely to find a high dust density at a
    location where the azimuthal gas velocity is high.}
  \label{zonalcomp}
\end{figure*}

In isothermal geostrophic balance, $2 \rho \Omega u_y = \cs^2 \dpa \rho / \dpa
x$, the azimuthal gas velocity follows the radial density gradient. That this
is true for large scales as shown in \Fig{zonalcomp}. The upper left panel
shows the evolution of the azimuthally and vertically averaged azimuthal
component of the gas velocity. Overplotted are the locations of the maxima in
the dust density. In the upper right panel the dust density evolution of the
same \run{L} is plotted. In comparing the location and times of the maxima and
minima on these two plots, one clearly sees that maxima in the dust density
occur often at times and locations where one finds maxima in the gas
velocity. Two attempts to quantify this observation are shown in the lower row
of \Fig{zonalcomp}. In the left panel, the particle density and the azimuthal
velocity from the two upper panels are plotted against each other, regardless
of position and time. In the right panel, a snapshot of the simulation (as
in \Fig{surfcoll}) was taken at $85$ local orbits, the time when the maximum
dust density enhancement occurs. The particle density as well as the azimuthal
gas velocity were integrated in vertical direction and plotted against each
other, regardless of their radial or azimuthal position in the simulation, in
this scatter plot. In order to visualize high densities of points in these
plots, we computed a two-dimensional histogram of the scattered points. This
is indicated by the color scale, showing the amount of points in each of the
boxes in the scatter plot space. There is a clear trend for high dust density
concentrations to appear at high gas velocities. Without radial drift
particles would concentrate where $u_y = 0$, i.e., between the sub- and
super-Keplerian flow. Due to the radial drift particles accumulate slightly
downstream at the formed pressure bumps. Those happen to be at the maxima of
the azimuthal gas velocity. With the geostrophic balance, high velocities are
also regions of a high radial density gradient. These plots prove that the
particles in the simulations are trapped by the long-lived pressure gradients
that occur due to stable zonal flows.

If the dust-to-gas ratio increases to values larger than unity, the streaming
instability \citep{YG05,JY07,YJ07} is triggered. This increases the dust
density further on timescales shorter than an orbital period. To follow the
streaming instability development, the back-reaction of the dust particles to
the gas phase must be considered in future numerical simulation. This effect
was neglected in this set of simulations. Otherwise the initial dust-to-gas
ratio would have been an additional free parameter to be studied.

\subsection{Radial Drift}\label{drift}

\begin{figure}[htb]
  \centering
    \includegraphics[width=\linewidth]{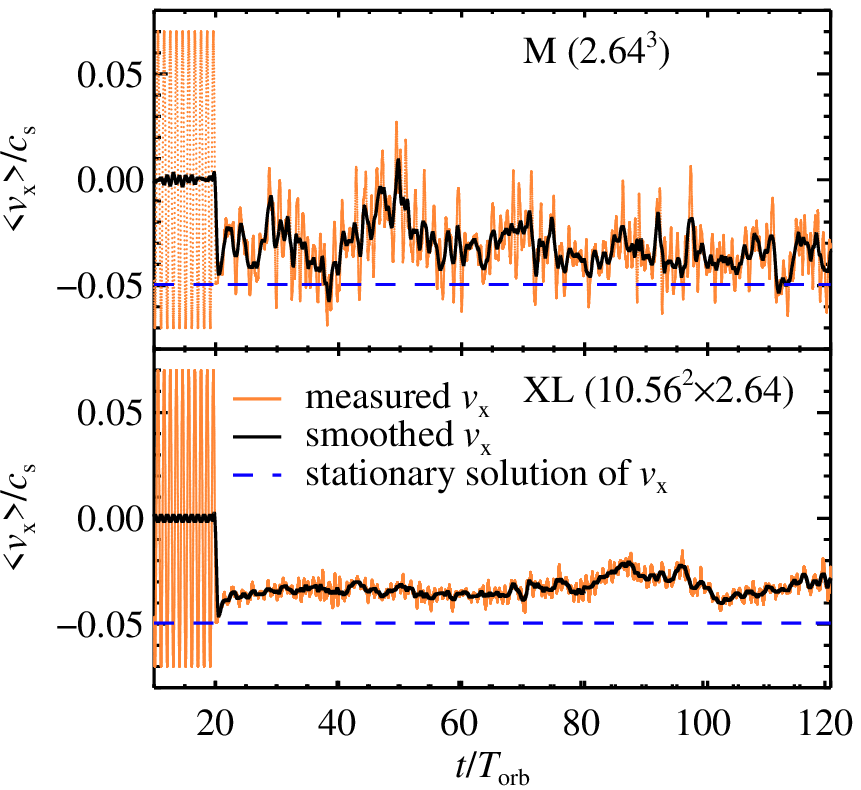}
    \includegraphics[width=\linewidth]{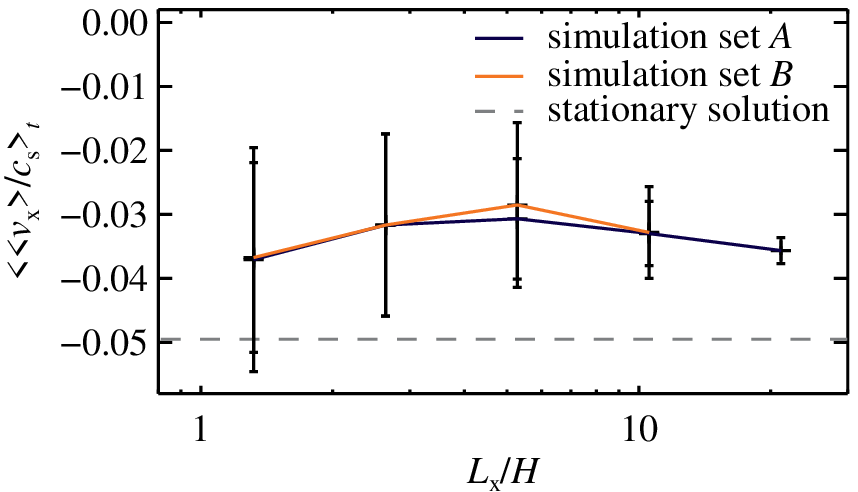}
  \caption{Radial drift velocity of the particles for two different
    simulations is shown in the upper panel. Particles in these simulations
    all have a Stokes number of $\St = \ts \Omega = 1$. The orange lines are
    the exact measured radial velocities, averaged over all particles. The
    black line represents the same value smoothed over the time of one local
    orbit. The blue dashed line shows the analytical result in a stationary
    box for particles of $\St = 1$ following \Eq{partvelinit}. Particles in
    turbulent simulations generally drift slower than expected from the
    stationary solution, but the box size has little effect on the drift
    velocity. This is shown in the lower panel where the mean of the radial
    drift velocity is plotted against the box size. We omitted simulation set
    \textit{C} for visibility. There is a minimum in drift speed for
    \run{L}. However, this minimum is within the error margins. The smallest
    errors are with \run{XXL}; here the drift velocity drops by $28\%$.}
  \label{radialvp}
\end{figure}

Radial drift velocities of the particles in the simulations with different box
sizes are shown in \Fig{radialvp}. The upper panel shows the measured and
expected radial drift of two simulations (\textit{M} and \textit{XL}). They show
that particles drift slower in turbulent simulations than they would in a
laminar disk. However, the size of the simulation has little effect on the
actual drift velocity, as shown in the lower panel of \Fig{radialvp}. It shows
a time average of the particle drift velocity plotted against the box
size. The uncertainties are too large to reveal a trend. Thus, the reduction
of the radial drift velocity apparently only depends on the amplitude of the
zonal flow, but not on the correlation time. Looking at the largest \run{XXL},
we can estimate that the radial drift gets reduced by about $28\%$ (drop of
the absolute value from $0.05 \cs$ to $(0.036 \pm 0.003)\cs$).

\begin{figure}[htb]
  \centering
    \includegraphics[width=\linewidth]{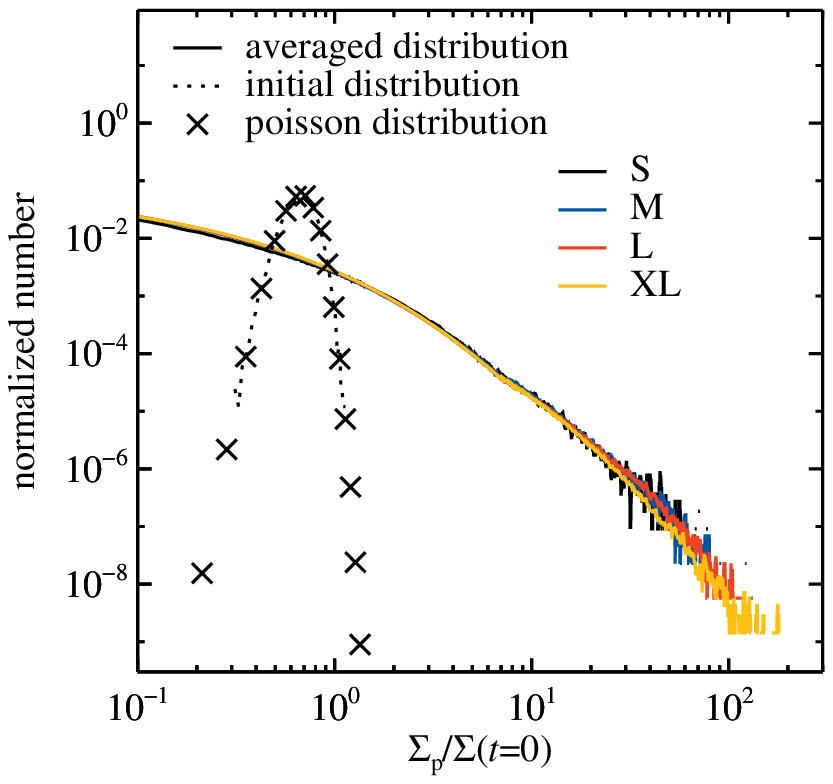}
  \caption{Distribution of the dust-to-gas ratio of the surface densities
    for \runsss{S}{M}{L}{XL}. For comparison a Poisson distribution is shown
    with crosses. The initial distribution of the numerical simulation (dotted
    line) fits very well to the normal distribution. The average strength of
    the clustering for the dust surface density does not depend on the
    simulation box size.}
  \label{clustering}
\end{figure}

\subsection{Clustering}\label{chapclustering}

The clustering degree of the particle distribution can be estimated with the
distribution of the dust surface density $\Sigma_p$ \citep{PPSKN11}. The initial
distribution is represented by a Poisson distribution (see
\Fig{clustering}).\footnote{Run~\textit{XXL} was not included in this figure,
  because the number of particles per grid cell was different to the other
  runs.} For this plot, we binned the measured dust surface density of a
snapshot. We then normalized them to the amount of grid cells. About three
local orbits after the particles feel the gas drag, the shape of the
distribution function is saturated. We averaged the distribution over the time
of $23 \ldots 121 \Torb$. We see at the high density end of the distribution
that higher densities develop in larger boxes due to the higher number of
available particles. Thus, the clustering properties do not depend strongly on
the strength or lifetime of the zonal flows (compare \Fig{rhopmax}, bottom).

\subsection{Different Particle Sizes}\label{chapspecies}

So far we only considered simulations with one particle species, i.e.,
$\St = \ts \Omega = 1$. We take the simulation size that simulates one fully
extended zonal flow and investigate $12$ different particle species. The
particle sizes range from $\St = 0.01$ to $\St = 100$. We choose \run{L} with
the dimensions $5.28H \times 5.28H \times 2.64H$ as simulation size for the
last simulation set. For one simulation we used a smaller box, because the
integration time had to be increased be a factor of two to give the particles
with the high Stokes numbers the opportunity to react on the pressure
differences.

\subsubsection{Drift Velocity and Particle Densities}\label{chapparden}

\begin{figure*}[t!]
  \centering
    \includegraphics[width=0.49\linewidth]{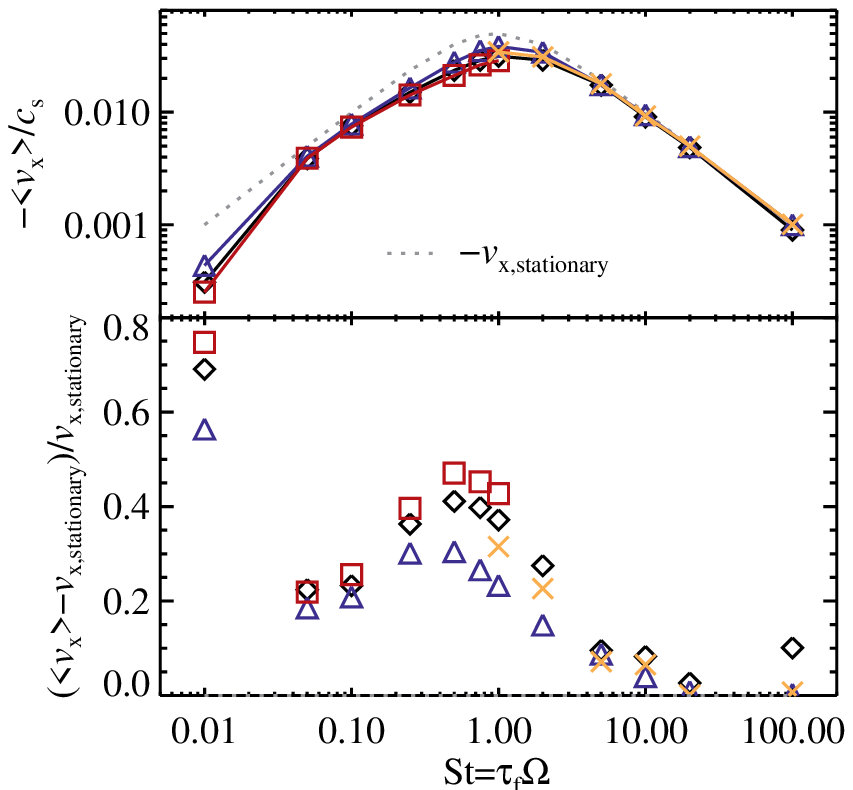}
    \includegraphics[width=0.49\linewidth]{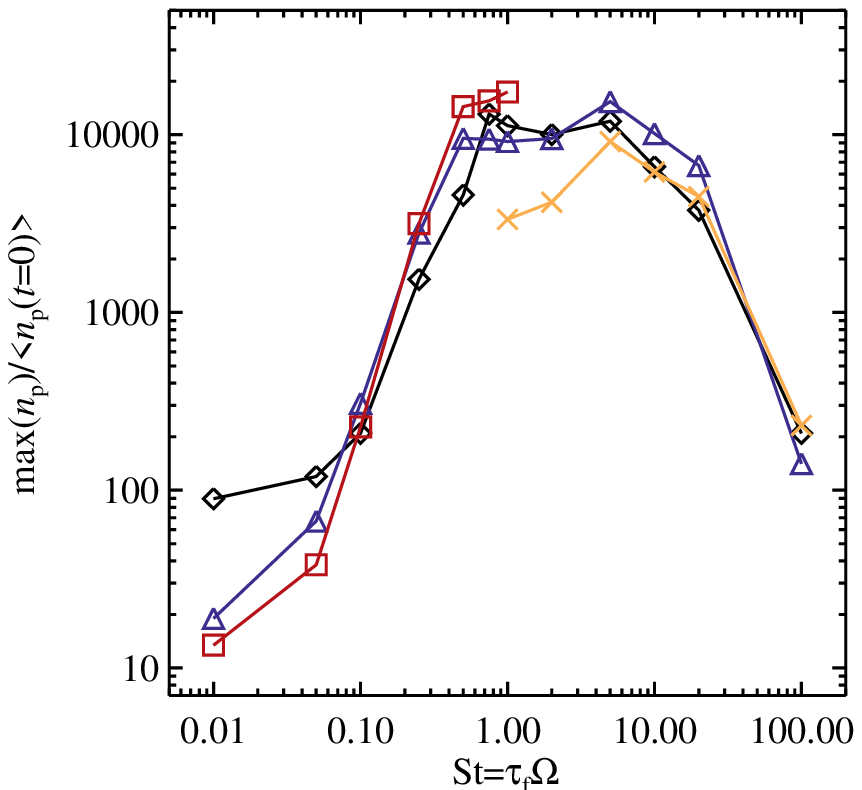}
    \includegraphics[width=0.49\linewidth]{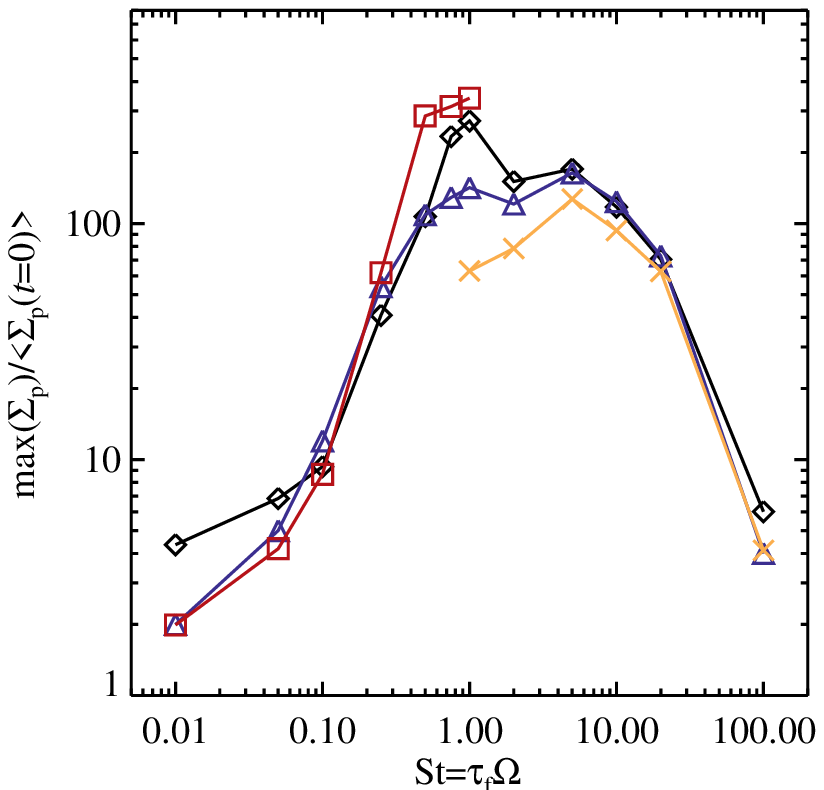}
    \includegraphics[width=0.49\linewidth]{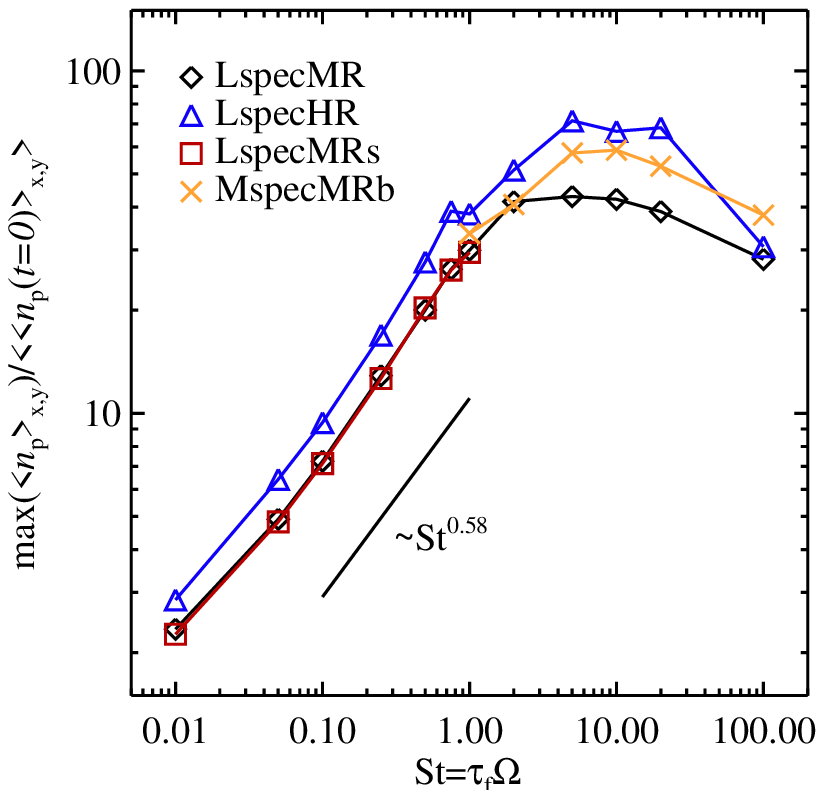}
  \caption{These panels show the behavior of particles with Stokes numbers
    from $0.01$ to $100$. The upper left panel shows the negative of the
    radial drift velocity and the relative difference between the measured and
    expected drift velocity. The dashed gray line shows the stationary
    solution for the radial drift, following \Eq{partvelinit}. The highest
    drift velocities are obtained for particles with $\St = 1$, but they are
    also slowed down strongest by the MRI-turbulence. The upper right panel
    shows the highest overdensity that occurred for the specific particle
    size during the entire simulation. The slopes for the different
    simulations match very well, apart from a jump around $\St = 1$ (for
    \textit{LspecMRs} and \textit{MspecMRb}) and an offset for \run{LspecMR}
    at small particle sizes. The former can be explained with the usage of a
    smaller simulation box for \run{MspecMRb} ($2.64^3$ with weaker zonal
    flows) than in the other simulations ($5.28^2 \times 2.64$ with stronger
    zonal flows). The offset showed that the number of particles per particle
    size was not sufficient in \run{LspecMR} ($10^5$ particles in $\sim 1.5
    \times 10^6$ grid cells leads with five particles in one grid cell to a
    result of $\max{(n_{\textrm{p}})}/\langle n_{\textrm{p}} (t = 0) \rangle =
    75$). The lower left panel shows the maximum of the column density for
    each particle size. It peaks at sizes of around $\St = 1$. The lower right
    panel shows the maxima of the vertical distribution of particles. The
    curves (for $\St = 0.01 \ldots 1$) follow a power-law with the index of
    $0.58 \pm 0.03$. This is slightly steeper than the expected power law
    index of $0.5$ \citep{DMS95}. In all four plots, the results of particles
    with $\St = 0.01$ are to be interpreted with caution, because the
    simulations lacked sufficient amount of super-particles for these size
    bins. Further, large particles ($\St = 100$) did not have enough time to
    sediment to the mid-plane.}
  \label{species}
\end{figure*}

The results are shown in \Fig{species}. The upper left panel shows the
negative of the radial velocity of the particles, averaged over all particles
of a certain size and over time. The four different simulations match very
well. The plot shows that particles with $\St = 1$ drift fastest inward, also
with turbulence in the simulations. On both sides the inward drift velocity
decreases with similar slopes. The key to the different colors and symbols is
in the lower right panel. Overplotted, in a dashed gray line, we find the
analytical prediction (following \Eq{partvelinit}) for the radial drift in a
laminar disk. The difference to the prediction is shown in the lower
sub-panel. Large particles generally drift slower according to the
steady-state solution and their coupling to the gas is also much weaker. Hence
their radial drift velocity is almost not affected by the turbulence and they
do not show strong concentrations. Small particles with low Stokes numbers
are stronger coupled to the gas and, thus, also drift very slow. Particles with
$\St \sim 1$ are concentrated most by the zonal flow and, thus, have a
stronger decreased radial velocity. Thus, the accumulation of dust particles
is expected to be strongest for particles with Stokes numbers around
unity. For $\St = 0.01$ particles, the drift velocity is strongly determined
by the gas flow. This explains the strong deviation from the expected drift
velocity.

The upper right panel shows the total particles overdensity normalized to the
initial particle number density. For \run{LspecMR} (black diamonds), the
smallest particles have higher concentrations than in the other
simulations. This resulted from the choice of too few particles per grid
cell. There only $100$,$000$ particles per size bin were simulated. This
results in overestimation, because the number density is normalized with the
initial number density $n_0$. For example, \run{LspecMRs} (red squares)
follows $2$,$000$,$000$ particles per particle size bin. The highest
concentrations were reached for particles of sizes $\St = 0.75 \ldots 5$, as
expected. However, the exact peak has a stochastic factor to it. Thus, the
simulations peak at different particle sizes. The overdensities are more
investigated in the lower row of panels.

The surface number density of the particles is shown in the lower left
panel. Here, the particles were integrated in the vertical direction. The
trend is similar to the upper right panel. We read from this plot that
particles with $\St = 0.1$ are concentrated about ten times the initial
concentration. Together with the vertical overdensity due to sedimentation
(lower right panel), a total overdensity of about $100$ is created for
$\St = 0.1$ particles.

The peaks in the vertical density structure of the particles are shown in the
lower right panel of \Fig{species}. The Stokes number, $\St = \ts \Omega$
defines the timescale after which the particles are settled down to the
mid-plane. Particles with a high Stokes number are not fully settled down to
the mid- plane, not even in the long-integration \run{MspecMRb}. The resolution
also limits this measurement for particles that are very close to the
mid-plane. Smaller particles are not that strongly stratified. Thus, the
vertical (Gaussian) structure is wider and shallower. This results in a lower
value in this plot. The points for Stokes numbers $0.01$--$1$ follow a power
law with the index of $0.58 \pm 0.03$. The measured power law index is
slightly higher than the expected value of $0.5$ \citep{DMS95}. Most of the
particles with $\St \gtrsim 1$ sediment very close to the mid-plane. This
prohibits a further increase in the vertical density. A higher resolution and
a measurement of the dust scale height is achieved in the next section. 

\subsubsection{Dust Pressure Scale Height}\label{chapHD}

\begin{figure}[htb]
  \centering
    \includegraphics[width=\linewidth]{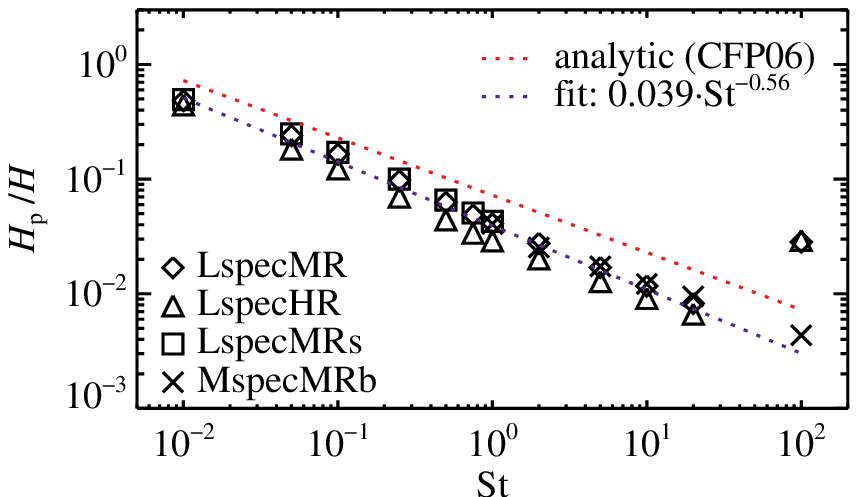}
    \includegraphics[width=\linewidth]{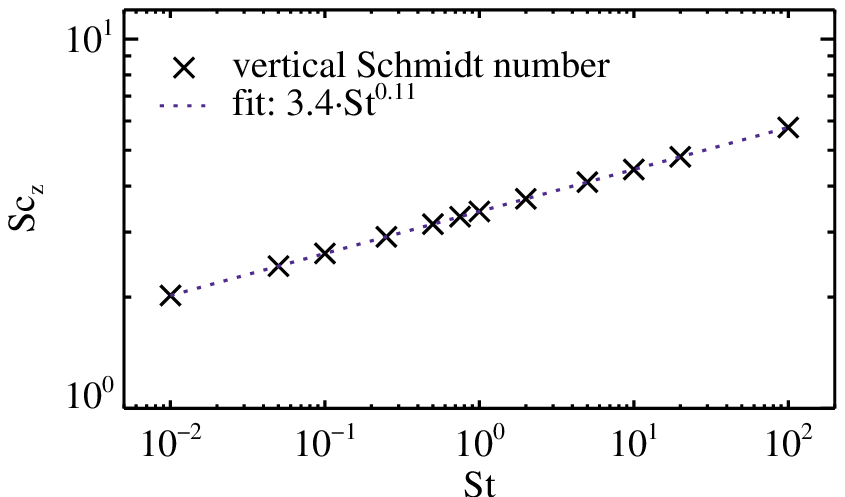}    
  \caption{Dust scale height as a function of the Stokes number is shown
    in the upper panel. Different symbols depict the different
    simulations. The expected dust scale height is calculated after
    \citet{CFP06} and compared with the fitted function. This value is the
    vertical Schmidt number $\textrm{Sc}_z$, shown in the lower panel. Its
    dependence on the particle size is very weak.}
  \label{dustscaleheight}
\end{figure}

With a stratified particle distribution we can test the vertical diffusion
model \citep[see, e.g.,][]{CFP06}. The dust pressure scale height can be
directly calculated from the vertical positions of the particles of the same
size. It is approximately proportional to $\St^{-0.5}$ in agreement with
\citet{CFP06,CBC11} and \citet{YL07}. The results are summarized in the upper
panel of \Fig{dustscaleheight}. Since the analytical value was calculated with
the $\alpha$-value, the vertical Schmidt number
\begin{equation}
  \textrm{Sc}_z = \frac{H_{\textrm{p, expected}}}{H_{\textrm{p, measured}}}
  \sim \left(\frac{\alpha}{D_T(\infty)} \right)^{\frac{1}{2}}
\end{equation}
can be calculated. We measured the vertical Schmidt number to have a very weak
dependence on the particle size. In the lower panel of \Fig{dustscaleheight}
we show that $\textrm{Sc}_z = 3.4 \cdot \St^{0.11}$.

\section{Discussion and Conclusions}\label{discussion}

\subsection{Zonal Flows and Axisymmetric Pressure Bumps}\label{zfdisc}

Our simulations have dimensionless units. This allows us to interpret our
results manyfold. We can pick the distance to the star in a certain range. In
\Sec{units}, we defined the global pressure gradient to be $\Delta v = 0.05
\cs$. In the minimum mass solar nebula (MMSN) model, we can choose the
distance to the star to be between $0.35$ and $40\,\textrm{AU}$
\citep{H81}. For this discussion, we pick $r = 5\,\textrm{AU}$. In a thin disk
model, we get a ratio for $H/r \sim\!0.033 (5\,\textrm{AU}/\textrm{AU})^{1/4}
\sim\!0.05$; this defines us $H = 0.25\,\textrm{AU}$. The isothermal sound
speed is $\cs = H \Omega \sim\!66$,$000\,\textrm{cm}~\textrm{s}^{-1}$. Thus,
turbulent velocities ($u_\textrm{rms}$) are about
$9000\,\textrm{cm}~\textrm{s}^{-1}$ ($\sim\!7000\,\textrm{cm}~\textrm{s}^{-1}$
for the high-resolution \run{LspecHR}).

\begin{figure}[htb]
  \centering
    \includegraphics[width=\linewidth]{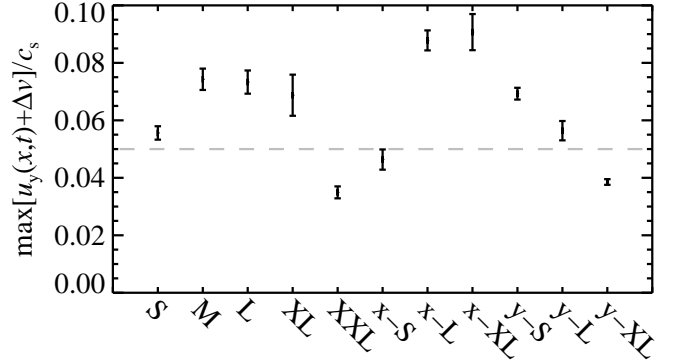}
  \caption{Measured highest azimuthal velocity of all box size simulated. The
    gray dashed line shows the threshold to Keplerian velocity. Only
    \runs{XXL}{y-XL} do never get super-Keplerian. Simulation~\textit{x-S}
    does get super-Keplerian at some time, but not often enough to be
    significant.}
  \label{superKepler}
\end{figure}

\Fig{superKepler} shows the highest azimuthal velocity for all simulation
sizes. We averaged over several maxima of $u_y(x,t)$ for every simulation to
smooth over outliers. The zonal flows are super-Keplerian for all but
\runss{XXL}{x-S}{y-XL}. In the largest box the flow only reaches slightly
sub-Keplerian velocities. However, particles still get captured in the
resulting axisymmetric pressure bumps. The speeds measured in the largest
simulation match those measured in \citet{FDKTH11}. 

We measured the radial size of the axisymmetric pressure bumps to be between
$5$ and $7 H$ (see \Fig{correlationtimes}). At a distance of $5\,\textrm{AU}$
to the star, this size corresponds to $\sim 1.25 \ldots 1.75 \,\textrm{AU}$
radial size for zonal flows, i.e., the distance between peaks of $\langle \rho
\rangle_{yz}$. This measurement agrees well with \citet{SBA12} who measured
the radial size of their zonal flows to be $6 H$. Further studies with varying
box size in smaller steps could potentially narrow down the radial scale.

We measured the lifetimes of the zonal flows up to $50 \Torb$. This agrees
well with earlier stated lifetimes \citep{JYK09,UKFH11}. The strength of the
density bump reaches $15\%$ and goes down to about $10\%$ in the largest
simulation. The lower amplitude is consistent with the results from global
simulations \citep[private communication with Mario Flock about the
simulations from][]{FDKTH11,FDKTH12} who measured a density enhancement of
slightly less than $10\%$. Some works \citep[e.g.,][]{UKFH11,SBA12} measure
stronger density enhancements. A possible explanation is that their $\alpha$
values are higher than in this work. Further studies on the dependence of
volume average quantities to strength of zonal flows would be interesting.

\subsection{Dust in Zonal Flows}\label{dustdisc}

Particles get trapped downstream of pressure bumps and build up
overdensities. To compare our dimensionless particle sizes with
collision experiments and observations we have to assume a distance to the
star and pick a solar system model. This will allow us to discuss our results
in context to recent experiments.

By choosing a model for the solar system, we can convert the dimensionless
Stokes number $\St = \ts \Omega$ to a real particle size. The friction time
$\ts$ correlates to the particle radius $a$ with 
\begin{equation}
  a = \frac{\ts^{\left( \textrm{Ep} \right)} \Omega \Sigma_{\textrm{gas}}}{\sqrt{2\pi}
    \rho_\bullet} \, , \label{taufEp} 
\end {equation}
for Epstein drag and
\begin{equation}
  a = \sqrt{\frac{9 \ts^{\left( \textrm{St} \right)} \Omega \mu H}{4 \rho_\bullet
      \sigma_\textrm{mol}}} \, , \label{taufSt} 
\end{equation}
for Stokes drag \citep[see supplementary info for][]{JOMKHY07}. Here
$\Sigma_{\textrm{gas}}$ is the column density of the gas, $\rho_\bullet$ is
the density of solid material, $\mu=3.9 \times 10^{-24}\,\textrm{g}$ is the
mean molecular weight, and $\sigma_{\textrm{mol}}=2 \times
10^{-15}\,\textrm{cm}^2$ is the molecular cross section of molecular hydrogen
\citep{NSH86,CC70}.

The Epstein regime applies, if the particle radius $a$ does not exceed $(9/4)$
\citep{W77a} of the gas mean-free path
\begin{equation}
  \lambda = \frac{\mu}{\rho_{\textrm{gas}} \sigma_{\textrm{mol}}} =
  \frac{\sqrt{2 \pi} \mu H}{\Sigma_{\textrm{gas}} \sigma_{\textrm{mol}}} \,
  . \label{gasmfp}
\end{equation}

\begin{figure}[htb]
  \centering
    \includegraphics[width=\linewidth]{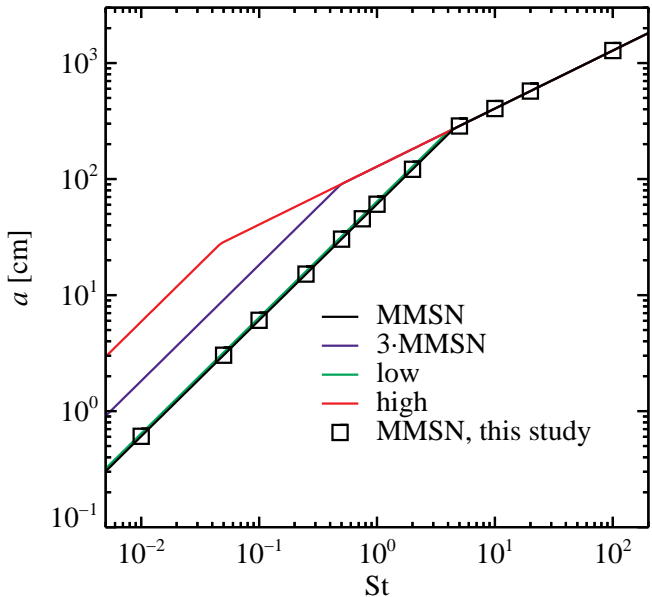}
  \caption{Particle sizes as a function of the dimensionless Stokes number for
    the four discussed models at $5\,\textrm{AU}$. The black squares show the
    used Stokes numbers and their corresponding size in the case of the
    minimum mass solar nebula model.}
  \label{parsizes}
\end{figure}

The gas density and hence also the particle size for a given Stokes number 
$\St = \ts \Omega$ depends very much on the used model. In \Fig{parsizes} we
overview four different models. The MMSN \citep{W77b,H81} was calculated from
the mass of the existing planets, neglecting migration. Because this model
allows no mass loss through accretion, often $3 \cdot$MMSN is used to account
for some accretion. A low-density model was published by \citet{BDH08}. This
model is adopted from measurements that indicate a shallow surface density
profile for protoplanetary disks \citep{AWHQD10}. The high-density model was
adopted from \citet{D07}, who introduced a ``revised MMSN model'' by using
the starting positions in the Nice model of planetary dynamics
\citep{TGML05}. This model also takes planetary migration into account. The
equations used to calculate the particle sizes in \Fig{parsizes} are
\begin{equation} \label{eqnmodels}
  \Sigma_{\textrm{gas}} = 
  \begin{cases}
    1700 \frac{\textrm{g}}{\textrm{cm}^2} \left( \frac{r}{\textrm{AU}}
    \right)^{-1.5} \, \textrm{(MMSN)} \\ 
    5100 \frac{\textrm{g}}{\textrm{cm}^2} \left( \frac{r}{\textrm{AU}}
    \right)^{-1.5} \, (3 \cdot \textrm{MMSN)} \\
    683 \frac{\textrm{g}}{\textrm{cm}^2} \left( \frac{r}{\textrm{AU}}
    \right)^{-0.9} \, \textrm{(low~density)} \\
    51\textrm{,}000 \frac{\textrm{g}}{\textrm{cm}^2} \left(
      \frac{r}{\textrm{AU}} \right)^{-2.2} \, \textrm{(high~density).} \\
  \end{cases}
\end{equation}
Throughout the discussion, we assume the MMSN model at $5\,\textrm{AU}$
distance to the star for size reference for our test particles. This choice
affects only the translation from the Stokes number $\St$ to a size, not the
dynamics in our models.

If the local dust density exceeds the Roche density, a clump is
gravitationally bound against shear. The Roche density can be approximated
\citep{K89} by
\begin{eqnarray} \label{eqnroche}
  \rho_{\textrm{Roche}}(R &=& 5\,\textrm{AU}) = \frac{9}{4 \pi}
  \frac{\Omega^2}{G(R=5\,\textrm{AU})} \nonumber \\
  &\sim& 100 \rho(R=5\,\textrm{AU}) \, ,
\end{eqnarray}
for an MMSN. $G$ is the gravitational constant. The streaming instability
\citep{YJ07,JY07} starts to act at dust-to-gas ratios of order unity. We
started all our simulations with $\epsilon_0 = \rho_p(t = 0) / \rho =
0.01$. Thus, a concentration of $\max(n_p)/\langle n_0 \rangle = 100$
corresponds to $\epsilon_{\textrm{streaming}} = 1$. The Roche density at
$5\,\textrm{AU}$ in an MMSN can be expressed as $\epsilon_{\textrm{Roche}} =
\rho_{\textrm{Roche}} / \rho \sim 100$. We can see that objects of several
decimeters up to some meters reach $\epsilon_{\textrm{Roche}}$, while pebbles
of some centimeters up to a decimeter reach $\epsilon_{\textrm{streaming}}$
from combining \Figs{species}{parsizes}. The concentration factors of
\run{LspecHR} in the upper right panel of \Fig{species} show us that with an
initial dust-to-gas ratio of $\epsilon_0 = 10^{-2}$ particles of sizes $\St =
0.5 \ldots 10$ ($\St = 0.1 \ldots 0.25$) reach a dust-to-gas ratio of $100$
($\gtrsim 1$). These sizes translate to $30 \ldots 400\,\textrm{cm}$ ($6
\ldots 15\,\textrm{cm}$) in an MMSN at a $5\,\textrm{AU}$ orbit using
\Fig{parsizes}. Considering back-reaction from the dust to the gas would allow
the streaming instability to act. This will be subject of a future study. In
our simulations, we see that the density of $15 \ldots 600\,\textrm{cm}$ sized
icy boulders increases several thousand times over the equilibrium density,
even without streaming instability and self-gravity of the
particles. Sedimentation to the mid-plane leads to overdensities of around
$40$, while the contribution from the turbulence concentrates the boulders
several hundred times.

Since we do not study the influence of the back-reaction from particles to the
gas, we were able to study several particle sizes in one simulation. That also
means that the initial dust-to-gas ratio ($\epsilon_0$) can be set
arbitrary. We can interpret our results in the light of different
metallicities. Particles with $\St \geq 0.5$ will trigger the streaming
instability even with $\epsilon_0 = 10^{-4}$, while $\St = 0.1$ particles need
$\epsilon_0 = 10^{-2}$.

At the assumed distance in this discussion, the resulting rings of trapped
dust are not observable with current telescopes. If zonal flows form at larger
distances to the star and dust rings form at an observable size, they could
potentially be observable with ALMA. For an analysis one would have to adjust
the parameter $\Delta v$ to account for the steeper pressure gradient. A
preliminary study showed that particles of about $10\,\textrm{cm}$ in size can
get capture for a short amount of time at $100\,\textrm{AU}$
distance. However, this question goes beyond the scope of this paper and
should be addressed in a future study.

\section{Summary and Outlook}\label{summary}

We performed numerical simulations of MRI-driven turbulence in shearing boxes,
covering the parameter space for radial and azimuthal box sizes up to
$21.12H$. Further, we followed the reaction of the dust particle density
to the turbulence. Our major findings are as follows.
\begin{enumerate}
  \item{Turbulent energy and stresses double when increasing the azimuthal
      size of the simulation from $1.32$ to $2.64$ pressure scale heights.
      Turbulence parameters in radially small box sizes stay approximately
      constant. This confirms the results in \citet{FS09}. In larger boxes,
      turbulent fluctuations and stresses are observed to remain constant
      against changes in the box size \citep[see also][]{JYK09}. This rapid
      convergence was also observed in \citet{SBA12}.}
  \item{Surface density fluctuations grow to large scales in the box and have
      lifetimes of up to $50$ orbits. The scales of these pressure bumps
      increase with increasing radial box size, until it saturates at
      approximately $5$--$7$ pressure scale heights. The scales are
      decreased when the azimuthal box size is much more increased than the
      radial box size. The radial scales of the pressure bumps are consistent
      with the length scales measured in local \citep[e.g.,][]{JYK09,SBA12}
      and global \citep[e.g.,][]{LJKP08,UKFH11} simulations. This might be the
      natural size of these overdensities. The pressure bumps are in
      geostrophic balance with sub- and super-Keplerian zonal flows. At
      $5\,\textrm{AU}$ distance to the star $6H$ correspond to $\sim
      1.5\,\textrm{AU}$. The amplitude of the density bump reaches $15\%$ and
      goes down to about $10\%$ in the largest simulation.}
  \item{Particles with $\St = \ts \Omega = 1$ are getting trapped efficiently
      by the axisymmetric pressure bumps. They accumulate in regions of minima
      in the second derivative of the gas density as predicted analytically
      \citep[e.g.,][]{KL01}. The concentration factor correlates with the
      correlation time of the zonal flows. Hence, the first two steps of
      planetesimal formation\footnote{After coagulation from $\mu$m-sized
        particles to $\St=0.1,1$.} in protoplanetary disk with an acting MRI
      are: vertical settling via sedimentation and radial concentration by
      trapping of dust in axisymmetric pressure bumps. Further concentration
      comes likely from stochastic processes. Clustering properties do not
      depend strongly on strength or lifetime of the zonal flows.}
  \item{We reach dust-to-gas ratios of $50$--$100$. These densities are of
      the order of the Roche density at $5\,\textrm{AU}$ in an MMSN. The dust
      overdensities scale with the lifetime of the zonal flow structures by
      a power law with an exponent of $0.38 \pm 0.05$ (see \Fig{rhopmax}).
      To what degree these high dust-to-gas ratios disturb the axisymmetric
      pressure bumps that developed in the zonal flows has to be
      investigated in further studies with back-reaction to the gas.}
  \item{Particles of only a few centimeters in size (at $5\,\textrm{AU}$ in an
      MMSN, $\St = 0.1$) accumulate in overdensities that are increased by a
      factor of $\sim 100$, leading to a dust-to-gas ratio of $1$ in the
      mid-plane, thus triggering the streaming instability. Without MRI and
      zonal flows $\St = 0.1$ particles do not clump strongly and cannot
      trigger the streaming instability for solar metallicity $Z = \epsilon_0
      = 0.01$ \citep{JYM09}.}
\end{enumerate}

This is the first work on the effect from large-scale zonal flows on dust
particles in an MHD simulation. Dust gets trapped downstream of long-lived
high-pressure regions and achieves overdensities that have the potential to
generate streaming instability and to become gravitationally
unstable. Planetesimal formation in large boxes will be further investigated
in simulations with particle feedback on the gas and self-gravitating
particles in a future study.

In the future, we will focus on one model and study various initial
dust-to-gas ratios and particle size distributions. We will probably use the
already converged \run{L} ($5.28H \times 5.28H \times 2.64H$). This choice is
also a trade-off between simulation box size and computational expense.

\begin{acknowledgements}

K.D. and H.K. have been supported by the Deutsche Forschungsgemeinschaft
Schwerpunktprogramm (DFG SPP) 1385 ``The first ten million years of the solar
system''; K.D. received further support by the IMPRS for Astronomy \& Cosmic
Physics at the University of Heidelberg. A.J. was partially funded by the
European Research Council under ERC Starting Grant agreement
278675–PEBBLE2PLANET and by the Swedish Research Council (grant
2010–3710). Computer simulations have been performed at the THEO cluster at
Rechenzentrum Garching and at the JUGENE cluster in 
Forschungszentrum J\"{u}lich (project number HHD19). We thank an anonymous
referee for a very thorough referee report that improved the quality of the
paper greatly.

\end{acknowledgements}


\begin{thebibliography}{35}

\bibitem[Andrews et al.(2010)]{AWHQD10} Andrews, S.~M., Wilner, D.~J., Hughes,
  A.~M., Qi, C., \& Dullemond, C.~P.\ 2010, \apj, 723, 1241
\bibitem[Balbus \& Hawley(1991)]{BH91} Balbus, S.~A., \& Hawley, J.~F.\ 1991,
  \apj, 376, 214
\bibitem[Balbus \& Hawley(1998)]{BH98} Balbus, S.~A., \& Hawley, J.~F.\ 1998,
  RvMP, 70, 1 
\bibitem[Beitz et al.(2011)]{BGBMTW11} Beitz, E., G{\"u}ttler, C., Blum, J.,
  et al.\ 2011, \apj, 736, 34
\bibitem[Birnstiel et al.(2012)]{BKE12} Birnstiel, T., Klahr, H., \& Ercolano,
  B.\ 2012, \aap, 539, A148  
\bibitem[Blum \& Wurm(2008)]{BW08} Blum, J., \& Wurm, G.\ 2008, \araa,
  46, 21 
\bibitem[Brandenburg et al.(1995)]{BNST95} Brandenburg, A., Nordlund, A.,
  Stein, R.~F., \& Torkelsson, U.\ 1995, \apj, 446, 741 
\bibitem[Brauer et al.(2008)]{BDH08} Brauer, F., Dullemond, C.~P., \& Henning,
  T.\ 2008, \aap, 480, 859  
\bibitem[Carballido et al.(2011)]{CBC11} Carballido, A., Bai, X.-N., \& Cuzzi,
  J.~N.\ 2011, \mnras, 415, 93
\bibitem[Carballido et al.(2006)]{CFP06} Carballido, A., Fromang, S., \&
  Papaloizou, J.\ 2006, \mnras, 373, 1633
\bibitem[Cassen \& Moosman(1981)]{CM81} Cassen, P., \& Moosman, A.\ 1981,
  Icar, 48, 353  
\bibitem[Chapman \& Cowling(1970)]{CC70} Chapman, S., \& Cowling, T.~G.\ 1970,
  The Mathematical Theory of Non-uniform Gases. An Account of the Kinetic
  Theory of Viscosity, Thermal Conduction and Diffusion in gases (3rd ed.;
  Cambridge: Univ. Press)   
\bibitem[Cuzzi et al.(2008)]{CHS08} Cuzzi, J.~N., Hogan, R.~C., \& Shariff,
  K.\ 2008, \apj, 687, 1432
\bibitem[Desch(2007)]{D07} Desch, S.~J.\ 2007, \apj, 671, 878  
\bibitem[Dominik et al.(2007)]{DBCW07} Dominik, C., Blum, J., 
  Cuzzi, J.~N., \& Wurm, G.\ 2007 in Protostars and Planets V ed. B. Reipurth,
  D. Jewitt, \& K. Keil (Tucson, AZ: Univ. Arizona Press), 783
\bibitem[Dubrulle et al.(1995)]{DMS95} Dubrulle, B., Morfill, G., \& Sterzik,
  M.\ 1995, Icar, 114, 237
\bibitem[Dzyurkevich et al.(2010)]{DFTKH10} Dzyurkevich, N., Flock, M.,
  Turner, N.~J., Klahr, H., \& Henning, T.\ 2010, \aap, 515, A70
\bibitem[Fedele et al.(2010)]{FAHJO10} Fedele, D., van den Ancker,
  M.~E., Henning, T., Jayawardhana, R., \& Oliveira, J.~M.\ 2010,
  \aap, 510, A72
\bibitem[Flock et al.(2011)]{FDKTH11} Flock, M., Dzyurkevich, N., Klahr, H.,
  Turner, N.~J., \& Henning, T.\ 2011, \apj, 735, 122
\bibitem[Flock et al.(2012)]{FDKTH12} Flock, M., Dzyurkevich, N., Klahr, H.,
  Turner, N., \& Henning, T.\ 2012, \apj, 744, 144 
\bibitem[Fromang \& Stone(2009)]{FS09} Fromang, S., \& Stone, J.~M.\ 2009,
  \aap, 507, 19
\bibitem[Guan et al.(2009)]{GGSJ09} Guan, X., Gammie, C.~F., Simon, J.~B., \&
  Johnson, B.~M.\ 2009, \apj, 694, 1010
\bibitem[Haisch et al.(2001)]{HLL01} Haisch, K.~E., Jr., Lada, E.~A., \& Lada,
  C.~J.\ 2001, ApJL, 553, 153 
\bibitem[Hayashi(1981)]{H81} Hayashi, C.\ 1981, PThPS, 70, 35
\bibitem[Heinemann \& Papaloizou(2009)]{HP09} Heinemann, T., \& Papaloizou,
  J.~C.~B.\ 2009, \mnras, 397, 64
\bibitem[Ida et al.(2008)]{IGM08} Ida, S., Guillot, T., \& Morbidelli, A.\
  2008, \apj, 686, 1292
\bibitem[Johansen et al.(2011)]{JKH11} Johansen, A., Klahr, H., \& Henning,
  T.\ 2011, \aap, 529, A62
\bibitem[Johansen et al.(2007)]{JOMKHY07} Johansen, A., Oishi, 
  J.~S., Mac Low, M.-M., et al.\ 2007, Natur, 448, 1022 
\bibitem[Johansen \& Youdin(2007)]{JY07} Johansen, A., \& Youdin, A.\ 2007,
  \apj, 662, 627
\bibitem[Johansen et al.(2009a)]{JYK09} Johansen, A., Youdin, A., \& Klahr, H.\
  2009a, \apj, 697, 1269
\bibitem[Johansen et al.(2009b)]{JYM09} Johansen, A., Youdin, A., \& Mac Low,
  M.-M.\ 2009b, ApJL, 704, 75
\bibitem[Klahr \& Lin(2001)]{KL01} Klahr, H.~H., \& Lin, D.~N.~C.\ 2001, \apj,
  554, 1095 
\bibitem[Kopal(1989)]{K89} Kopal, Z.\ 1989, The Roche Problem and its
  Significance for Double-star Astronomy (Astrophysics and Space Science
  Library, Vol. 152; Dordrecht: Kluwer)
\bibitem[Krumholz et al.(2012)]{KKM12} Krumholz, M.~R., Klein, R.~I., \&
  McKee, C.~F.\ 2012, \apj, 754, 71
\bibitem[Lodders(2003)]{L03} Lodders, K.\ 2003, \apj, 591, 1220 
\bibitem[Lyra et al.(2008)]{LJKP08} Lyra, W., Johansen, A., Klahr, H., \&
  Piskunov, N.\ 2008, \aap, 479, 883 
\bibitem[Nakagawa et al.(1986)]{NSH86} Nakagawa, Y., Sekiya, M., \& Hayashi,
  C.\ 1986, Icar, 67, 375 
\bibitem[Pan et al.(2011)]{PPSKN11} Pan, L., Padoan, P., Scalo, J., Kritsuk,
  A.~G., \& Norman, M.~L.\ 2011, \apj, 740, 6  
\bibitem[Pinilla et al.(2012)]{PBRDUTN12} Pinilla, P., Birnstiel, T., Ricci,
  L., et al.\ 2012, \aap, 538, A114
\bibitem[Safronov(1969)]{S69} Safronov, V.~S. (ed.) 1969, Evoliutsiia
  doplanetnogo oblaka (English transl.: Evolution of the
  Protoplanetary Cloud and Formation of Earth and the Planets, NASA
  Tech. Transl. F-677, 1972,; Jerusalem: Israel Sci. Transl.)
\bibitem[Shakura \& Sunyaev(1973)]{SS73} Shakura, N.~I., \& Sunyaev, R.~A.\
  1973, \aap, 24, 337 
\bibitem[Simon et al.(2012)]{SBA12} Simon, J.~B., Beckwith, K., \& Armitage,
  P.~J.\ 2012, \mnras, 422, 2685
\bibitem[Sorathia et al.(2012)]{SRSB12} Sorathia, K.~A., Reynolds, C.~S.,
  Stone, J.~M., \& Beckwith, K.\ 2012, \apj, 749, 189
\bibitem[Stone \& Gardiner(2010)]{SG10} Stone, J.~M., \& Gardiner, T.~A.\
  2010, \apjs, 189, 142
\bibitem[Tsiganis et al.(2005)]{TGML05} Tsiganis, K., Gomes, R., Morbidelli,
  A., \& Levison, H.~F.\ 2005, Natur, 435, 459  
\bibitem[Uribe et al.(2011)]{UKFH11} Uribe, A.~L., Klahr, H., Flock, M., \&
  Henning, T.\ 2011, \apj, 736, 85
\bibitem[von Neumann \& Richtmyer(1950)]{vNR50} von Neumann, J., \& Richtmyer,
  R.~D.\ 1950, JAP, 21, 232
\bibitem[Weidenschilling(1977a)]{W77a} Weidenschilling, S.~J.\ 1977a,
  \mnras, 180, 57
\bibitem[Weidenschilling(1977b)]{W77b} Weidenschilling, S.~J.\ 1977b, \apss,
  51, 153
\bibitem[Weidenschilling(1997)]{W97} Weidenschilling, S.~J.\ 1997, Icar,
  127, 290
\bibitem[Whipple(1972)]{W72} Whipple, F.~L.\ 1972, in Proc. Twenty-First Nobel
  Symp. on From Plasma to Planet, ed. A. Evlius (New York: Wiley), 211
\bibitem[Windmark et al.(2012a)]{WBGBDH12} Windmark, F., Birnstiel, T.,
  G{\"u}ttler, C., et al.\ 2012a, \aap, 540, A73
\bibitem[Windmark et al.(2012b)]{WBOD12} Windmark, F., Birnstiel, T., Ormel,
  C.~W., \& Dullemond, C.~P.\ 2012b, \aap, 544, L16
\bibitem[Youdin \& Johansen(2007)]{YJ07} Youdin, A., \& Johansen, A.\ 2007,
  \apj, 662, 613
\bibitem[Youdin \& Goodman(2005)]{YG05} Youdin, A.~N., \& Goodman, J.\ 2005,
  \apj, 620, 459
\bibitem[Youdin \& Lithwick(2007)]{YL07} Youdin, A.~N., \& Lithwick, Y.\ 2007,
  Icar, 192, 588
\bibitem[Zsom et al.(2010)]{ZOGBD10} Zsom, A., Ormel, C.~W., G{\"u}ttler, C.,
  Blum, J., \& Dullemond, C.~P.\ 2010, \aap, 513, A57

\expandafter\ifx\csname natexlab\endcsname\relax\def\natexlab#1{#1}\fi
\end{thebibliography}
\end{document}